\documentclass[twocolumn,trackchanges]{aastex701}
\usepackage{enumitem}
\usepackage{amsmath}
\usepackage{booktabs}
\usepackage{subcaption}
\hypersetup{hypertexnames=false}

\begin{document}

\title{The Sun’s chemical peculiarity: disentangling Galactic chemical evolution and planetary engulfment in solar twins}

\author[orcid=0009-0006-7542-9639]{Mia Babatsikos}
\affiliation{School of Physics and Astronomy, Monash University, Clayton, Victoria 3800, Australia.}
\affiliation{OzGrav: Australian Research Council Centre of Excellence for Gravitational Wave Discovery, Clayton, VIC 3800, Australia.}
\email[show]{mia.babatsikos@monash.edu}  

\author[orcid=0000-0003-4794-6074]{Fan Liu} 
\affiliation{National Astronomical Observatories, Chinese Academy of Sciences, Beijing 100101, China.}
\affiliation{School of Physics and Astronomy, Monash University, Clayton, Victoria 3800, Australia.}
\email[show]{fanliu@bao.ac.cn}

\author[orcid=0000-0002-3625-6951]{Amanda Karakas}
\affiliation{School of Physics and Astronomy, Monash University, Clayton, Victoria 3800, Australia.}
\email{amanda.karakas@monash.edu}

\author[orcid=0000-0002-6134-8946]{Ilya Mandel}
\affiliation{School of Physics and Astronomy, Monash University, Clayton, Victoria 3800, Australia.}
\affiliation{OzGrav: Australian Research Council Centre of Excellence for Gravitational Wave Discovery, Clayton, VIC 3800, Australia.}
\email{ilya.mandel@monash.edu}

\author[orcid=0009-0000-2982-2217]{Lachlan Passenger}
\affiliation{School of Physics and Astronomy, Monash University, Clayton, Victoria 3800, Australia.}
\affiliation{OzGrav: Australian Research Council Centre of Excellence for Gravitational Wave Discovery, Clayton, VIC 3800, Australia.}
\email{lachlan.passenger@monash.edu}

\author[orcid=0000-0001-9907-7742]{Megan Bedell}
\affiliation{Center for Computational Astrophysics, Flatiron Institute, 162 5th Avenue, New York, NY 10010, USA.}
\email{mbedell@flatironinstitute.org}

\author[orcid=0000-0002-6937-9034]{Sharon X.~Wang}
\affiliation{Department of Astronomy, Tsinghua University, Beijing 100084, China.}
\email{sharonw@tsinghua.edu.cn}

\author[orcid=0009-0006-4529-6724]{Zimo Cheng}
\affiliation{Department of Astronomy, Tsinghua University, Beijing 100084, China.}
\email{zimocheng1119@gmail.com}

\begin{abstract}
Recent observational studies have suggested that the Sun may be chemically peculiar relative to the majority of solar twins. Here, we re-analyse high-resolution, high signal-to-noise spectra of 79 nearby solar twins using a differential spectroscopic approach and Bayesian framework to test whether the Sun's chemical peculiarity arises from Galactic chemical evolution (GCE) or planetary ingestion. Using the spectroscopic tool \texttt{Korg}, we obtain highly precise, validated atmospheric parameters and abundances for 18 elements, with an average abundance precision of 0.015\,dex (3.5\%). Employing an independent Bayesian indicator, we disentangle GCE and planetary engulfment signatures from other processes influencing stellar composition, including intrinsic abundance scatter. Our results indicate that the chemical peculiarity of the Sun relative to the average solar twin is largely driven by GCE effects, with 62.3$\pm$5.8\% of our sample exhibiting abundance patterns well-described by GCE trends. We further identify 2--6 solar twin candidates exhibiting chemical signatures consistent with planetary engulfment that warrant further investigation. These findings reinforce the importance of accounting for GCE effects when interpreting solar twin abundance patterns, and suggest that the Sun may not be chemically peculiar relative to the majority of solar twins.
\end{abstract}



\section{Introduction}\label{sec:introduction}
Spectroscopic observations of solar-type main-sequence stars provide a powerful means of placing our Solar System in a broader Galactic context and exploring the link between stars and their planetary systems. Solar twins are stars with solar-like atmospheric parameters including effective temperature $T_{\mathrm{eff}}$, surface gravity $\log g$, microturbulence $\xi$, and metallicity (taken here as iron abundance [Fe/H]). They are particularly valuable observational targets because their near-identical photospheres to the Sun enable highly precise, line-by-line differential abundance analyses (\citealp{gustafsson_2025} and references therein). This approach effectively removes systematic biases implicitly encoded in both the solar and solar twin abundance patterns through subtraction, including those arising from stellar atmosphere modelling uncertainties \citep{asplund_2005}. As a result, solar twin abundance uncertainties can be reduced below the level of 0.01\,dex (2\%) \citep{melendez_2009, melendez_2014, bedell_2014, bedell_2018}. 

Studies of solar twins are crucial for assessing whether the Sun and its planets are typical. Through analysis of 11 solar twins, \cite{melendez_2009} found evidence suggesting that the Sun may have a peculiar abundance pattern relative to an average sample of solar twins. They found the Sun is depleted in refractory elements relative to volatiles by 20\% compared to the average solar twin composition, quantified by a trend in elemental abundances with condensation temperature ($T_{\rm cond}$). The discovery of this trend sparked debate over the Sun’s chemical peculiarity, with subsequent work yielding mixed results (e.g., \citealp{ramirez_accurate_2009, ramirez_2010, gonzalez_2010, gonzalez_hernandez_2010, gonzalez_hernandez_2013, bedell_2018}). A key limitation of this approach is its reliance on $T_{\rm cond}$ values, which depend sensitively on the condensation environment (see Section \ref{sec:discussion}) and therefore introduce additional uncertainties into the inferred abundance trends \citep{spaargaren_2025}.

The physical origin of this solar anomaly remains under debate due to the numerous processes that may affect solar twin abundance patterns. Solar twin compositions are influenced by Galactic chemical evolution (GCE) effects correlated with the star's time and place of birth in the Galaxy. However, recent studies suggest the solar $T_{\rm cond}$ trend may persist even after accounting for GCE effects, though results are sensitive to the adopted assumptions \citep{adibekyan_2014, nissen_2015, nissen_2016, adibekyan_2016, bedell_2018, cowley_galactic_2022, carlos_peculiar_2025, martos_2025, rampalli_galactic_2026}. Additional processes such as atomic diffusion, planet formation, magnetic activity, gas-dust segregation in the protoplanetary disk, and dust cleansing in the primordial nebula are also expected to affect the photospheric composition of stars. The nature of these effects remains an area of ongoing investigation \citep{chambers_2010,onehag_2014, gaidos_2015, dotter_2017,booth_owen_2020, yu_c3po_2025}.

Planet ingestion resulting from dynamical instabilities during the main sequence provides an alternative explanation for the solar abundance anomaly (e.g., \citealp{pinsonneault_2001, sevilla_2022}). The small outer convective zones expected during this evolutionary stage allow accreted material to potentially leave a strong and long-lasting effect on photospheric composition. In this scenario, the average solar twin engulfs rocky planet material and becomes enriched in refractories, while in contrast, our Sun may have avoided such events, leaving it relatively depleted. This theory implies that a history of strong dynamical interactions is common in Sun-like systems.

The origin of the peculiar solar composition therefore remains uncertain, with previous studies limited by their reliance on $T_{\rm cond}$ values and linear fitting of abundance trends. In this work, we re-analyse the solar twin sample presented by \cite{bedell_2018} and \cite{spina_2018} using the new spectroscopic analysis tool \texttt{Korg} \citep{korg, korg2} and apply Bayesian inference techniques to characterise the observed abundance trends. This implementation of \texttt{Korg} provides an opportunity to test the capability of the code and facilitates independent verification of the stellar parameters and abundances reported by \cite{spina_2018} and \cite{bedell_2018}. Furthermore, the Bayesian framework adopted is informed by the possible physical origins of the solar peculiarity. This method enables interpretations that are independent of $T_{\rm cond}$ trends and more physically motivated than the linear fitting processes implemented in previous studies. Using this framework, we investigate whether the chemical patterns observed in solar twins are better explained by GCE effects or planetary ingestion.

While binary stellar twin systems are often favoured for studying planet engulfment because their co-natal evolution minimises systematic abundance differences, our work instead focuses on non-co-natal field solar twins. Observations of solar-type binaries have revealed abundance signatures consistent with main-sequence planet ingestion, implying occurrence rates of $\sim$2--10\% \citep{behmard_2023a, liu_2024}, in agreement with current theoretical models \citep{behmard_2023a, oconnor_lai_2025}. By extending this investigation to field solar twins, we aim to place constraints on the occurrence rates of planetary engulfment in the broader population of Sun-like stars.

In Section \ref{sec:method}, we describe our spectroscopic analysis method and implementation of an independent Bayesian indicator. In Section \ref{sec:results}, we present the derived stellar parameters and abundances, and identify stars with compositions well-described by GCE effects, as well as those showing evidence of planet engulfment. We discuss the implications of our findings in Section \ref{sec:discussion} and provide concluding remarks in Section \ref{sec:conclusion}.

\section{Method} \label{sec:method}

\subsection{Data}
The solar twin sample used in this study consists of 79 solar twins within 100\,pc and is adopted from previous solar twin studies by \cite{bedell_2018} and \cite{spina_2018}. This sample was curated using selection criteria requiring a stellar effective temperature within 100\,K of solar; surface gravity within 0.1\,dex of solar; and metallicity within 0.1\,dex of solar. 

High-resolution stellar spectra for these solar twins were obtained using the High Accuracy Radial velocity Planet Searcher (HARPS) spectrograph mounted on the 3.6\,m telescope at La Silla Observatory in Chile \citep{mayor_2003}. Additional spectral data from the MIKE spectrograph were incorporated to 
supplement the oxygen dataset, as the infrared coverage required to measure the O\,I triplet lying at wavelengths 7771.944, 7774.166 and 7775.388\,Å was not available in the HARPS spectra \citep{bernstein_2003, ramirez_2013}. These spectra possess both high resolving power (HARPS: R = 115,000; MIKE: R = 83,000) and high typical signal-to-noise ratios (HARPS: $\sim$800 per pixel at 600\,nm; MIKE: $\sim$400 per pixel at 600\,nm). Finally, the solar reference spectrum was produced by combining multiple exposures of sunlight reflected by the asteroid Vesta, with a signal-to-noise ratio of approximately 1300 per pixel at 600\,nm \citep{bedell_2018}. 

The equivalent widths employed in this work have been extracted from the subsequent analyses of these spectra conducted by \cite{bedell_2018} and \cite{spina_2018}. Further information regarding the reduction processing of the spectra and equivalent width measurements can be found in these studies. 

\subsection{Stellar Parameters and Abundances}\label{subsec:stellar_params_and_abunds}
We re-analysed the solar twin sample using a differential, line-by-line equivalent width technique implemented with the recently-developed \texttt{Korg} 1D LTE spectral analysis tool \citep{korg, korg2}, in conjunction with MARCS model atmospheres \citep{marcs}. In contrast, \cite{spina_2018} and \cite{bedell_2018} analysed the same stellar sample using the \texttt{MOOG} 1D LTE spectral analysis tool \citep{moog} together with Kurucz ATLAS9 model atmospheres \citep{castelli_kurucz_2004}. The advantage of using \texttt{Korg} is its increased calculation speed (up to 100 times faster than similar spectroscopic codes), and incorporation of improved physics for calculating chemical equilibrium and model equivalent widths \citep{korg, korg2}. These improvements include updated continuum opacities and partition functions, a more modern molecular/electron equation of state, and refined treatment of hydrogen lines.

The stellar parameters reported in this study were derived using the \texttt{Fit.ews\_to\_stellar\_parameters} function within \texttt{Korg}. This method achieves best-fitting stellar parameters by evaluating the excitation-ionisation balance, using Fe\,I and Fe\,II equivalent width measurements with interpolations from MARCS model atmospheres. A Newton-Raphson solver is used to achieve the closest possible excitation-ionisation balance to zero. The final stellar parameters were then obtained by applying global offsets of $T_{\rm eff}$ = +64.6\,K, log\,$g$ = +0.006\,dex, $\xi$ = -0.015\,$\rm km\,s^{-1}$, and [Fe/H] = +0.103 across the entire stellar sample to ensure consistency with the standard solar values: $T_{\rm eff}$ = 5777\,K, log\,$g$ = +4.44\,dex, $\xi$ = 1\,$\rm km\,s^{-1}$, and [Fe/H] = 0 (e.g., \citealp{cox_2000}). Following this normalisation process, the differential excitation-ionisation balances remain sufficiently close to zero across the stellar sample.

Line-by-line differential stellar abundances for 18 elements (C, O, Na, Mg, Al, Si, S, Ca, Sc, Ti, V, Cr, Mn, Fe, Co, Ni, Cu, and Zn) were computed for all 79 stars using the \texttt{Korg} function \texttt{Fit.ews\_to\_abundances}, initialised with [$\alpha$/Fe] = 0. Different elemental ionisation states were treated as separate species. Hyperfine structure has not been taken into account owing to incompatibility with the \texttt{Korg} linelist, duplicate carbon lines were consolidated by averaging, and outlier spectral lines ($>2\sigma$ from the species mean) were omitted. Nevertheless, the agreement between our derived abundances and those of \cite{bedell_2018} (shown in Section \ref{subsec:stellar_param_abunds_results}) indicates these adjustments do not significantly affect our results, particularly given the differential nature of the analysis within a strictly defined sample of solar twins. Non-local thermodynamic equilibrium (NLTE) corrections were interpolated from \cite{amarsi_2015} and applied to the O\,I abundances. 

Final differential abundances for each species were obtained by subtracting the calculated solar abundance from the corresponding solar twin abundance on a line-by-line basis, and then averaging over all lines (i.e., taking the geometric mean). Uncertainties on both the parameters and abundances were calculated using the methodology outlined in \cite{bensby_2014} and \cite{liu_2014}, which accounts for both line-to-line scatter and the influence of stellar parameter variations on the abundance change.

To reduce the influence of atomic diffusion on the solar twin abundance patterns, we primarily employ [X/Fe] abundances, as opposed to [X/H]. Atomic diffusion alters the surface composition when chemical species sink and rise under the competing processes of radiative acceleration and gravitational settling \citep{dotter_2017}. While this process affects iron in a similar manner to other metals - causing gradual diffusion out of the photosphere with increasing stellar age - hydrogen exhibits the opposite trend, increasing in photospheric concentration with time. This behaviour is theoretically entangled in the [X/H] abundances, but can be minimised by normalisation via subtraction of [Fe/H] \citep{bedell_2018}, mitigating age-dependent atomic diffusion effects on a star-by-star basis. Bayesian modelling of atomic diffusion was not employed, as the available models \citep{dotter_2017} cover only eight of the 18 elements analysed in this work. Nevertheless, the relative effects of atomic diffusion between the Sun and our strictly defined solar twin sample are expected to be minimal, and thus should have little impact on the differential abundance patterns.

To remove GCE effects, we evaluated the \cite{bedell_2018} linear age--abundance relations at each star's age (using \citealt{spina_2018} estimates) to predict the expected [X/Fe] pattern, then subtracted this from the observed abundances. The resulting residuals encode how much each star's chemistry deviates from that of a typical star of its age in the solar neighbourhood, effectively isolating abundance variations unrelated to chemical evolution. Corresponding uncertainties on the GCE-corrected abundances for species $j$ and star $i$ were determined analytically by adding observational error $\sigma_{obs}$ in quadrature with the age-abundance trend uncertainties propagated to the abundance uncertainties, as shown in Equation (\ref{eq:GCE_uncert}). The GCE age-abundance slope and intercept uncertainties are $\sigma_m$ and $\sigma_b$ respectively, $m$ is the best fit of the GCE-age trend gradient from \cite{bedell_2018}, $age$ and $\sigma_{age}$ are the observed stellar age and corresponding measurement error from \cite{spina_2018}. 

\begin{equation}
\sigma [\mathrm{X_j/Fe}]_{i} = \sqrt{\sigma_{b,j}^2+\sigma_{m,j}^2*age_i^2+\sigma_{age, i}^2*m_j^2+\sigma_{obs, j,i}^2}
\label{eq:GCE_uncert}
\end{equation}

The GCE-corrected [X/Fe] abundances derived using the \cite{bedell_2018} trends form the primary dataset for our engulfment analysis, which we hereafter refer to as the \textit{GCE-corrected (B18)} sample. As a robustness check, we additionally construct a second set of GCE-corrected abundances using age-abundance trends derived directly from our own abundance measurements, hereafter referred to as the \textit{GCE-corrected (TW)} sample. The derivation of these trends is described in Appendix \ref{apdx:GCEageabundancetrends}. 

\subsection{Bayesian Method}\label{sec:bayesian_method}
We conduct a Bayesian analysis to examine the origin of the observed abundance patterns in our stellar sample, modelled after the approach of \cite{liu_2024}. While previous studies focusing on linear fitting have been limited by their use of $T_{\rm cond}$ trends to characterise solar twin abundance patterns, our Bayesian statistical technique is able to probe the physical origin of these patterns independent of $T_{\rm cond}$ values. 

The \texttt{dynesty} nested sampling algorithm \citep{dynesty} was 
applied to examine both the evidence and posterior probability 
simultaneously for the five scenarios described in Section 
\ref{subsubsec:bayesian_models}. For each star $i$, we adopted the 
standard Gaussian log-likelihood function of \cite{liu_2024}, which assumes that the observed abundance residuals follow a Gaussian distribution. 
We assume that the intrinsic abundance variation for each element is an independent random draw from a zero-centred Gaussian with standard deviation $\sigma_{\mathrm{scatter},i}$.  Marginalising over the abundance variations analytically yields the likelihood function
\begin{multline}
    \log \mathcal{L}([\mathrm{X/Fe}]_{\rm obs, i} | \theta) = {\rm norm} \\
    - \sum^{N}_{j}\frac{([\mathrm{X_j/Fe}]_{\rm obs,i} - [\mathrm{X_j/Fe}]_{\rm model,i})^2}{2(\sigma^2_{\mathrm{[X_j/Fe}],i} + \sigma^2_{\mathrm{scatter},i})}\ ,
\label{eq:logL}
\end{multline}
where $\sigma_{[\mathrm{X_j/Fe}],i}$ is the measurement uncertainty on the abundance of element $j$ in star $i$, $N$ is the number of elements and the logarithmic normalisation term (norm) is defined as: 
\begin{equation}
    {\rm norm} = - \frac{N \log(2\pi)}{2} - \sum^N_j \log \sqrt{(\sigma ^2 _{[\mathrm{X_j/Fe}],i} + \sigma^2_{scatter,i})}.
    \label{eq:norm}
\end{equation}
The scatter term is treated as a free parameter (see Section \ref{subsubsec:bayesian_models}) and marginalised over during nested sampling. 

Together, Equations (\ref{eq:logL}) and (\ref{eq:norm}) define the likelihood function $\mathcal{L}([\mathrm{X/Fe}]_{\rm obs,i} | \theta)$ used throughout this work. For each physical scenario considered, we compute the posterior distribution

\begin{equation}
P(\theta |[\mathrm{X/Fe}]_{\rm obs,i}) = \frac{\mathcal{L}([\mathrm{X/Fe}]_{\rm obs,i} | \theta) \, \pi(\theta)}{Z_i}
\end{equation}

where $\pi(\theta)$ is the prior distribution on the model parameters and $ Z_i = \int \mathcal{L}([\mathrm{X/Fe}]_{\rm obs,i} | \theta) \, \pi(\theta) \, \mathrm{d}\theta$ is the Bayesian evidence. The parameter space $\theta$ and choice of prior distributions $\pi(\theta)$ for each physical scenario are defined in Section \ref{subsubsec:bayesian_models}.

\subsubsection{Models}\label{subsubsec:bayesian_models}

We implemented two baseline models to serve as references for comparison with models incorporating astrophysical structure. Our `null-offset' model defines the baseline case where the stellar 
composition is identical to the Sun's ($[\mathrm{X/Fe}]_{\rm model} = 0$), 
such that any observed abundance deviations are attributed solely to intrinsic scatter $\sigma_{\rm scatter}$ - a free parameter representing 
the standard deviation of random abundance offsets around zero. For each abundance pattern we used model parameter $\theta \sim \{ \sigma_{scatter}\}$, with uniform prior distribution $\sigma_{scatter}\sim\mathcal{U}$(0, 0.1). To account for intrinsic abundance differences arising from slight variations in the primordial stellar compositions, the second model we implemented was a `flat' model, which assumes the solar twin abundance patterns can be explained by intrinsic abundance scatter ($\sigma_{scatter}$) and an overall abundance offset ($\delta$) such that the predicted abundance for a given star is $[\mathrm{X/Fe}]_{\rm model}$ = $\delta$. The abundance data was thus described using parameters $\theta \sim \{ \delta, \sigma_{scatter} \}$, with uniform prior distributions $\sigma_{scatter}\sim\mathcal{U}$(0, 0.1); $\delta \sim\mathcal{U}$(-0.3, 0.3).

For the models of planetary engulfment, we investigate the mass of planetary material required to be ingested into the solar twin convective zone in order to match the observed abundance patterns. These models assume the composition of an engulfed planet can be described as a mixture of bulk Earth composition and carbonaceous Mighei-like (CM) chondritic material, with elemental mass fractions $f_{X,BE}$ and $f_{X,CM}$ defined below following \cite{liu_2024}. CM chondritic material was chosen for this work as it represents the chemistry of the early Solar System \citep{deleuw_2010}, and has historically been used to represent volatile-rich planetary material as an alternative to volatile-poor bulk Earth material (e.g., \citealp{melendez_2009, chambers_2010, liu_2024}). The engulfment models also account for intrinsic scatter treated as a free parameter with the same prior as in the null-offset and flat model for consistency: $\sigma_{scatter} \sim\mathcal{U}(0, 0.1)$. The engulfment models are given by

\begin{equation}
[\mathrm{X/Fe}]_{\rm model} = [\mathrm{X/H}]_{\rm engulf} - [\mathrm{Fe/H}]_{\rm engulf},
\label{eq:[X/Fe]_abunds}
\end{equation}

\begin{equation*}
[\mathrm{X/H}]_{\rm engulf} = \log_{10} \left( \frac{f_{X, BE} M_{BE}+f_{X, CM} M_{CM}+M_{X, CZ}}{M_{X,CZ}} \right),
\label{eq:[X/H]_abunds}
\end{equation*}

and

\begin{equation}
M_{X, CZ} = \frac{10^{A(X)_{\odot}}m_X}{\sum_X 10^{A(X)_{\odot}} m_X} f_{CZ} M_{star},
\label{eq:M_X_CZ}
\end{equation}

where $M_{star}$ is the mass of the star, $f_{CZ}$ is the mass fraction of the stellar convection zone, $m_X$ is the atomic mass of element X, $M_{X, CZ}$ is the mass of element X already in the stellar convection zone, as determined by Equation (\ref{eq:M_X_CZ}), $f_{X, BE}$ and $f_{X, CM}$ are the elemental mass fractions of bulk Earth 
\citep{allegre_2001} and CM chondritic \citep{wasson_kallemeyn_1988} 
material respectively, $M_{BE}$ and $M_{CM}$ are respectively the masses of bulk Earth and CM chondritic composition material ingested into the stellar convection zone in this model. $A(X)_{\odot}$ is the abundance of species X in the Sun as derived by \cite{asplund_2009}.\footnote{$A(X) \equiv \log_{10}(N_X/N_H) + 12$, where $N_X$ and $N_H$ are the number densities of species X and hydrogen respectively.} For the stellar structure variables we assume solar values of $M_{star}=1\,M_{\bigodot}$ and $f_{CZ}=0.02$ \citep{christensen_dalsgaard_1991}, motivated by the strict selection criteria defining our solar twin sample (see Section \ref{subsec:limitations_futurework} for further discussion).

We use two different planet engulfment models distinguished by the composition of the ingested material. In the first model, the engulfed material is dominated by bulk Earth composition, with a small, fixed amount of CM chondritic material ($M_{CM} = 0.1\, M_{\oplus}$) to account for the possibility that an exoplanet's composition may not exactly match bulk Earth. The parameter set for this model is $\theta = \{ M_{BE}, \sigma_{scatter}\}$ 
with uniform prior distribution $M_{BE} \sim \mathcal{U}(1, 30) \, M_{\oplus}$.
In the second model, the engulfed material is dominated by CM chondritic 
composition, with a small, fixed amount of bulk Earth material 
($M_{BE} = 0.1 \, M_{\oplus}$). The parameter set for this model is 
$\theta = \{ M_{CM}, \sigma_{scatter}\}$ with uniform prior distribution 
$M_{CM} \sim \mathcal{U}(1, 30) \, M_{\oplus}$.

The last model implemented assumes solar twin abundance patterns are dominated by GCE effects. For a given star, the predicted abundance pattern is calculated using the empirically-derived \cite{bedell_2018} linear age-abundance relations for each species X, such that ${\rm[X/Fe]_{model}} = m_X \cdot age + b_X$. The slope and intercept values $m_X$ and $b_X$ are taken directly from \cite{bedell_2018}, while stellar age is a free parameter with a uniform prior $age \sim \mathcal{U}(0, 15)$\,Gyr. Although $m_X$ could in principle be treated as a free parameter, it is correlated with stellar age, and fitting both simultaneously would introduce a degeneracy. Our GCE model also accounts for intrinsic scatter as a free parameter, with the same priors as previous models for consistency. The parameter set for the GCE model is $\theta \sim \{age, \sigma_{scatter}\}$. 

Our analysis does not explicitly account for birth galactocentric radius. A star's chemical composition is primarily determined by the time and place of its birth, and observed abundance gradients along the Galactic disc - reproduced by GCE models (e.g., \citealp{matteucci_francois_1989, grisoni_2018}) - indicate that birth location is expected to contribute to the observed solar twin abundance patterns (see \citealp{martos_2025} and references therein). However, reliable determination of birth galactocentric radius is challenging and beyond the scope of this work; its potential effects are instead absorbed by our intrinsic scatter parameter, $\sigma_{scatter}$, along with other unresolved sources of abundance variation.

Uniform prior distributions were adopted for all model parameters to remain as uninformative as possible, with boundaries selected to encompass the observed data while allowing additional margin. For the parameters $\sigma_{scatter}$ and $\delta$, these limits were chosen based on the distribution of abundance measurements across all stars (Appendix~\ref{apdx:mocknoise_sample}). The prior boundaries on stellar age were selected from the age range of the sample derived by \cite{spina_2018}. For the engulfment mass priors, the lower bound was chosen to prevent degeneracy with the flat and null-offset models, while the upper bound was motivated by current understanding of refractory masses in planets. In the classical core accretion scenario, planets are expected to form from an initial rocky and icy core of up to 10\,$M_{\oplus}$ before accreting gas of nebular composition \citep{pollack_1996}. However, recent estimates of Jupiter’s composition suggest that giant planets may contain substantially larger masses of rock-forming elements, with Jupiter itself potentially hosting up to 40\,$M_{\oplus}$ of such material, although its bulk metallicity remains uncertain \citep{aguilera_gomez_2016, wahl_2017, guillot_2021, yildiz_2024}. Given these uncertainties, we adopted an upper prior limit of 30\,$M_{\oplus}$. Finally, we note that our implementation of mock-noise samples to calibrate the model-comparison criteria (Sections \ref{subsec:application_to_data} and \ref{sec:bayesian_results_overall}) should reduce the dependence of our final results on the choice of prior distributions.

\subsubsection{Application to Data}\label{subsec:application_to_data}
The models outlined in Section \ref{subsubsec:bayesian_models} were applied only to the 69 non-$\alpha$-enhanced solar twins in the sample (see Section \ref{subsec:stellar_param_abunds_results}). 

To understand the extent to which solar twin abundance patterns are shaped by GCE effects, we analysed the GCE-\textit{uncorrected} [X/Fe] abundance data using three models: null-offset, flat, and GCE. Conversely, potential planetary engulfment candidates were identified using GCE-\textit{corrected} [X/Fe] abundances, as these compositions are largely disentangled from age-dependent abundance trends (Section~\ref{subsec:stellar_params_and_abunds}). While models simultaneously incorporating both GCE and engulfment effects are possible in principle, our sequential approach of first correcting for GCE effects and then testing for engulfment is more interpretively transparent, albeit reliant on the accuracy of the adopted GCE trends and stellar age estimates. The GCE-corrected (B18) and GCE-corrected (TW) [X/Fe] abundances are analysed using four Bayesian frameworks: the null-offset and flat baseline models, together with the bulk Earth and CM chondritic engulfment models.

Our analysis focuses on planetary engulfment in the solar twin sample and does not consider the possibility that the Sun itself may have undergone engulfment. This choice is justified by the current stability of the Solar System architecture (see \citealp{lithwick_wu_2014} and references therein) and the observed refractory depletion of the Sun relative to solar twins (e.g., \citealp{melendez_2009}), which imply that recent planetary ingestion is unlikely.

To validate our results, we generated mock noise datasets representing our two baseline models: the null-offset and flat models. To create each mock realisation of the null-offset model, one of the 69 posterior distributions obtained by fitting the model to the observed data was randomly selected. A single value of the intrinsic scatter parameter ($\sigma$) was then bootstrap sampled from this posterior and adopted as the standard deviation of a zero-centred Gaussian distribution. From this distribution, 17 elemental abundance values were drawn according to $[\mathrm{X/Fe}]_{\rm noise} \sim\mathcal{N}(0, \sigma^2)$, consistent with the likelihood assumptions outlined in 
Section~\ref{sec:bayesian_method}. A corresponding set of 17 abundance uncertainties was bootstrap sampled from the empirical distribution of observed abundance uncertainties across all species and all non-$\alpha$-enhanced stars in the sample.

Mock noise for the flat model was generated in the same manner, except that a single value of the abundance offset parameter ($\delta$) was also bootstrap sampled from a randomly selected posterior distribution. The abundance values were then drawn from a Gaussian distribution centred on $\delta$ according to $[\mathrm{X/Fe}]_{\rm noise} \sim\mathcal{N}(\delta, \sigma^2)$. The abundance uncertainties were sampled identically to the null-offset case.

This procedure was repeated 1000 times for each baseline model (flat and null-offset), and across all three observed datasets -- the GCE-uncorrected and two GCE-corrected abundance datasets -- yielding 6000 mock samples in total. The resulting comparisons are presented in Appendix \ref{apdx:mocknoise_sample}.

\section{Results} \label{sec:results}
\subsection{Stellar Parameters and Abundances}\label{subsec:stellar_param_abunds_results}
The stellar parameters and abundances derived for our sample of 79 solar twins can be found in Tables \ref{tab:stellar_parameters} and \ref{tab:xh_abundances} respectively.

\begin{table*}
\centering
\caption{Atmospheric parameters for the solar twin sample.}
\label{tab:stellar_parameters}
\begin{tabular}{lcccc}
\hline\hline
Star &
$T_{\rm eff}$ &
$\log g$ &
$\xi$ &
$[{\rm Fe/H}]$ \\
 &
(K) &
(dex) &
(km\,s$^{-1}$) &
(dex) \\
\hline
HIP10175  & $5721.4 \pm 5.0$ & $4.49 \pm 0.012$ & $0.976 \pm 0.008$ & $-0.026 \pm 0.004$ \\
HIP101905 & $5902.6 \pm 5.2$ & $4.477 \pm 0.013$ & $1.07 \pm 0.008$ & $0.082 \pm 0.008$ \\
HIP102040 & $5839.2 \pm 5.1$ & $4.477 \pm 0.013$ & $1.012 \pm 0.008$ & $-0.088 \pm 0.005$ \\
HIP102152 & $5709.5 \pm 5.1$ & $4.336 \pm 0.013$ & $1.000 \pm 0.007$ & $-0.020 \pm 0.004$ \\
HIP10303  & $5707.9 \pm 4.1$ & $4.417 \pm 0.010$ & $0.969 \pm 0.007$ & $0.099 \pm 0.006$ \\
HIP104045 & $5820.5 \pm 4.3$ & $4.414 \pm 0.011$ & $1.039 \pm 0.007$ & $0.047 \pm 0.004$ \\
$\cdots$    & $\cdots$           & $\cdots $           & $\cdots$            & $\cdots$ \\
\hline
\end{tabular}
\begin{flushleft}
\textit{Note.} The full version of this table is available in electronic form.
\end{flushleft}
\label{table:stellar_params}
\end{table*}

\begin{table*}
\centering
\caption{Elemental abundances [X/H] for the solar twin sample.}
\label{tab:xh_abundances}
\begin{tabular}{lccccccc}
\hline\hline
Star &
[Al\,I/H] &
[C\,I/H] &
[Ca\,I/H] &
[Co\,I/H] &
$\cdots$ &
[Sc\,II/H] &
[Zn\,I/H] \\
 &
(dex) &
(dex) &
(dex) &
(dex) &
 &
(dex) &
(dex) \\
\hline
HIP10175  &
$-0.062 \pm 0.016$ &
$-0.059 \pm 0.016$ &
$-0.003 \pm 0.010$ &
$-0.060 \pm 0.016$ &
$\cdots$ &
$-0.047 \pm 0.012$ &
$-0.077 \pm 0.023$ \\

HIP101905 &
$+0.050 \pm 0.011$ &
$-0.062 \pm 0.049$ &
$+0.111 \pm 0.011$ &
$+0.031 \pm 0.021$ &
$\cdots$ &
$+0.068 \pm 0.011$ &
$-0.018 \pm 0.044$ \\

HIP102040 &
$-0.107 \pm 0.010$ &
$-0.100 \pm 0.032$ &
$-0.050 \pm 0.017$ &
$-0.125 \pm 0.020$ &
$\cdots$ &
$-0.058 \pm 0.023$ &
$-0.162 \pm 0.039$ \\

HIP102152 &
$+0.008 \pm 0.007$ &
$-0.004 \pm 0.012$ &
$-0.016 \pm 0.004$ &
$-0.030 \pm 0.009$ &
$\cdots$ &
$-0.015 \pm 0.004$ &
$-0.018 \pm 0.010$ \\

HIP10303  &
$+0.132 \pm 0.007$ &
$+0.093 \pm 0.015$ &
$+0.096 \pm 0.005$ &
$+0.138 \pm 0.020$ &
$\cdots$ &
$+0.124 \pm 0.009$ &
$+0.096 \pm 0.019$ \\

HIP104045 &
$+0.035 \pm 0.007$ &
$+0.006 \pm 0.017$ &
$+0.054 \pm 0.004$ &
$+0.034 \pm 0.008$ &
$\cdots$ &
$+0.046 \pm 0.007$ &
$+0.014 \pm 0.016$ \\

$\cdots$ & $\cdots$ & $\cdots$ & $\cdots$ & $\cdots$ & $\cdots$ & $\cdots$ & $\cdots$ \\
\hline
\end{tabular}

\begin{flushleft}
\textit{Note.} The full table, including all measured species, is available in electronic form.
\end{flushleft}
\label{table:stellar_abunds}
\end{table*}

\begin{figure}
    \centering
    \includegraphics[width=\linewidth]{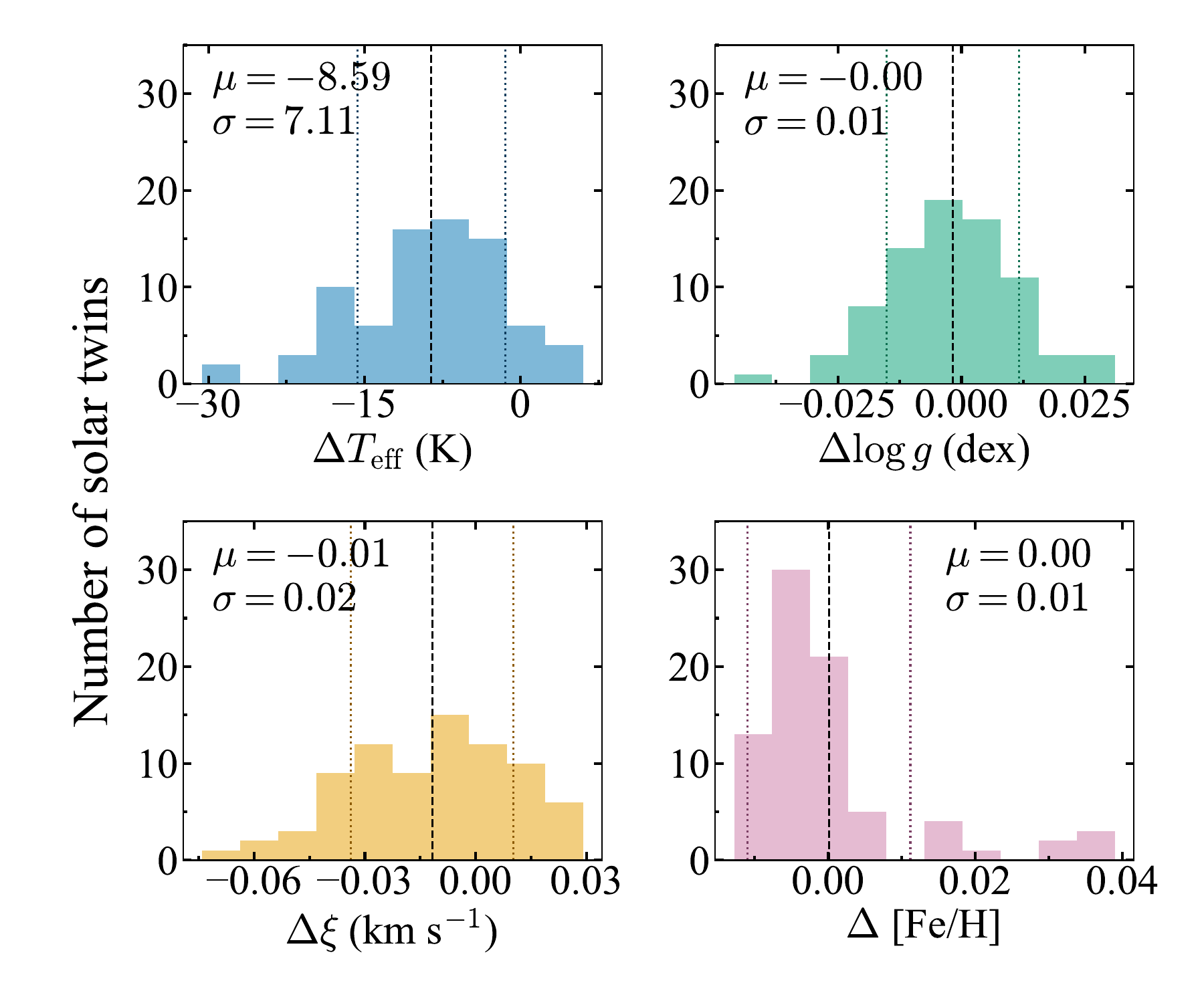}
    \caption{Distribution of differences in stellar parameters between this work and \cite{spina_2018} for the same sample of 79 solar twins, comparing effective temperature (top left), surface gravity (top right), microturbulence (bottom left) and metallicity (bottom right). The x-axis shows $\Delta m = m_{\rm this\ work} - m_{\rm Spina}$ for each star. Dashed and dotted lines indicate the mean $\mu$ and standard deviation $\sigma_{\rm diff}$ of each distribution respectively, with values quoted in each panel.}
    \label{fig:bedell_parameter_comparison}
\end{figure}

The typical stellar parameter uncertainties achieved are highly precise and comparable to precisions reported by \cite{spina_2018}, with average measurement error: $\sigma(T_{\rm eff})$=5.4\,K, $\sigma$(log\,$g$)=0.016\,dex, $\sigma (\xi)$=0.01\,$\rm km\,s^{-1}$, and $\sigma$([Fe/H])=0.008\,dex. Figure~\ref{fig:bedell_parameter_comparison} compares the stellar parameters derived in this work with those of \cite{spina_2018}. The star-to-star scatter in parameter differences (quantified by $\sigma_{\rm diff}$, the standard deviation of $\Delta m$ shown in Figure~\ref{fig:bedell_parameter_comparison}) is consistent with the combined measurement uncertainties for $T_{\rm eff}$ and $\log g$ ($\sigma_{\rm diff} /\sqrt{\langle\sigma_{\rm this\ work}^2 + \sigma_{\rm Spina}^2\rangle} \approx 0.98$ and $0.60$ respectively), while $\xi$ and $[\rm Fe/H]$ have slightly higher ratios of 1.36 and 1.12. Statistically significant mean offsets exist for $T_{\rm eff}$ and $\xi$, at $10.7\times$ and $4.8\times$ the standard error of the mean respectively, while $\log g$ and $[\rm Fe/H]$ show no significant systematic offset ($1.2\times$ and $0.1\times$ the standard error). 

To isolate the origin of these differences, we recomputed parameters using different combinations of code and model atmosphere. We find that changing the model atmosphere alone alters parameters negligibly, while changing the spectral analysis code produces small differences ($\lesssim$9\,K in $T_{\rm eff}$, $\lesssim$0.02\,dex in $\log g$, $\sim$0.01\,km\,s$^{-1}$ in $\xi$). This indicates that the differences with \cite{spina_2018} discussed above are primarily driven by the choice of spectral analysis code rather than the model atmosphere.

\begin{figure}
    \centering
    \includegraphics[width=1\linewidth]{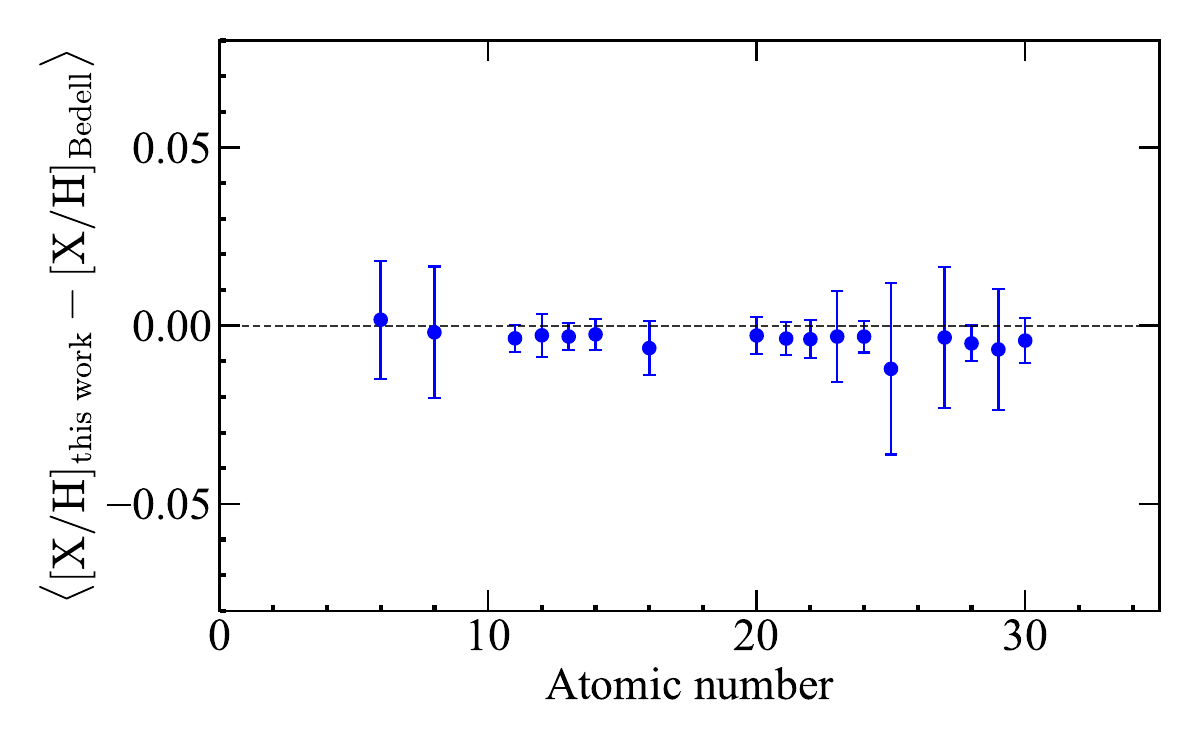}
    \caption{Mean differences in differential elemental abundances [X/H] between this work and \cite{bedell_2018}, averaged over the stellar sample of 79 solar twins, as a function of atomic number. Error bars represent the standard deviation of the abundance differences across the stellar sample ($\sigma_{\rm diff}$). The average abundance difference between the two studies is -0.004\,dex.}
    \label{fig:abundance_comparison_with_bedell}
\end{figure}

We achieve a high average abundance precision of 0.015\,dex (3.5\%), the same order of magnitude as \cite{bedell_2018}. Figure~\ref{fig:abundance_comparison_with_bedell} presents the mean abundance differences between this work and \cite{bedell_2018} for individual species. Statistically significant mean offsets exist for 13 of 17 species, at between $2.1\times$ and $8.8\times$ the standard error of the mean, with magnitudes typically $\lesssim$0.01\,dex. These offsets primarily reflect differences between the \texttt{Korg} and \texttt{MOOG} spectral analysis codes, with a smaller contribution from the choice of model atmosphere (typical differences of $\sim$0.002–0.004\,dex from model atmosphere compared to $\sim$0.007\,dex from code). Nevertheless, we find the star-to-star scatter in abundance differences is smaller than the combined measurement uncertainties ($\sigma_{\rm diff} < \sqrt{\langle \sigma_{\rm this\ work}^2 + \sigma_{\rm Bedell}^2\rangle}$) for all 17 species, confirming excellent star-to-star consistency between the two studies. Additionally, analysis of our derived abundances recovers the same 10 solar twins identified as $\alpha$-enhanced by \cite{bedell_2018} using the same criteria; these stars are older than 8\,Gyr and possess a visible enhancement in $\alpha$ elements ([$\alpha$/Fe]$>1\sigma$ based on our \texttt{Korg}-derived abundance data). These $\alpha$-enhanced stars deviate from the linear age-abundance GCE trends outlined by \cite{bedell_2018}, and are therefore not considered further in this study.

Overall the close star-to-star agreement between the results obtained using \texttt{Korg} and \texttt{MOOG} for an identical equivalent width dataset demonstrates the reliability of \texttt{Korg}. This high precision combined with fast speed (discussed in Section \ref{subsec:stellar_params_and_abunds}) highlights \texttt{Korg} as a competitive tool for differential analysis in future large spectroscopic surveys. These results also demonstrate that differential analysis substantially suppresses systematic errors that arise from varying physical implementations of the excitation-ionisation balance and different model atmospheres.

\subsection{Bayesian Results}\label{sec:bayesian_results_overall}
We obtain Bayesian evidence values quantifying the ability of each model to describe the observed abundance patterns. Model selection was performed by inspecting the Bayesian evidence difference $\Delta \ln Z = \ln Z_{\rm model\,1} - \ln Z_{\rm model\,2}$, also known as the natural log Bayes factor, which measures the relative statistical preference for one model over another based on their ability to reproduce the abundance patterns. To assess the significance of these model preferences, the same Bayesian framework was applied to the mock noise realisations described in Section \ref{subsec:application_to_data}. The resulting mock-noise $\Delta \ln Z$ distributions were then used to estimate probability values ($p$-values), representing the probability -- under the baseline or null-hypothesis models -- of obtaining evidence differences at least as extreme as those observed.

\subsubsection{Galactic Chemical Evolution Effects}\label{sec:GCE_bayesian_results}
\begin{figure}
    \centering
    \includegraphics[width=1\linewidth]{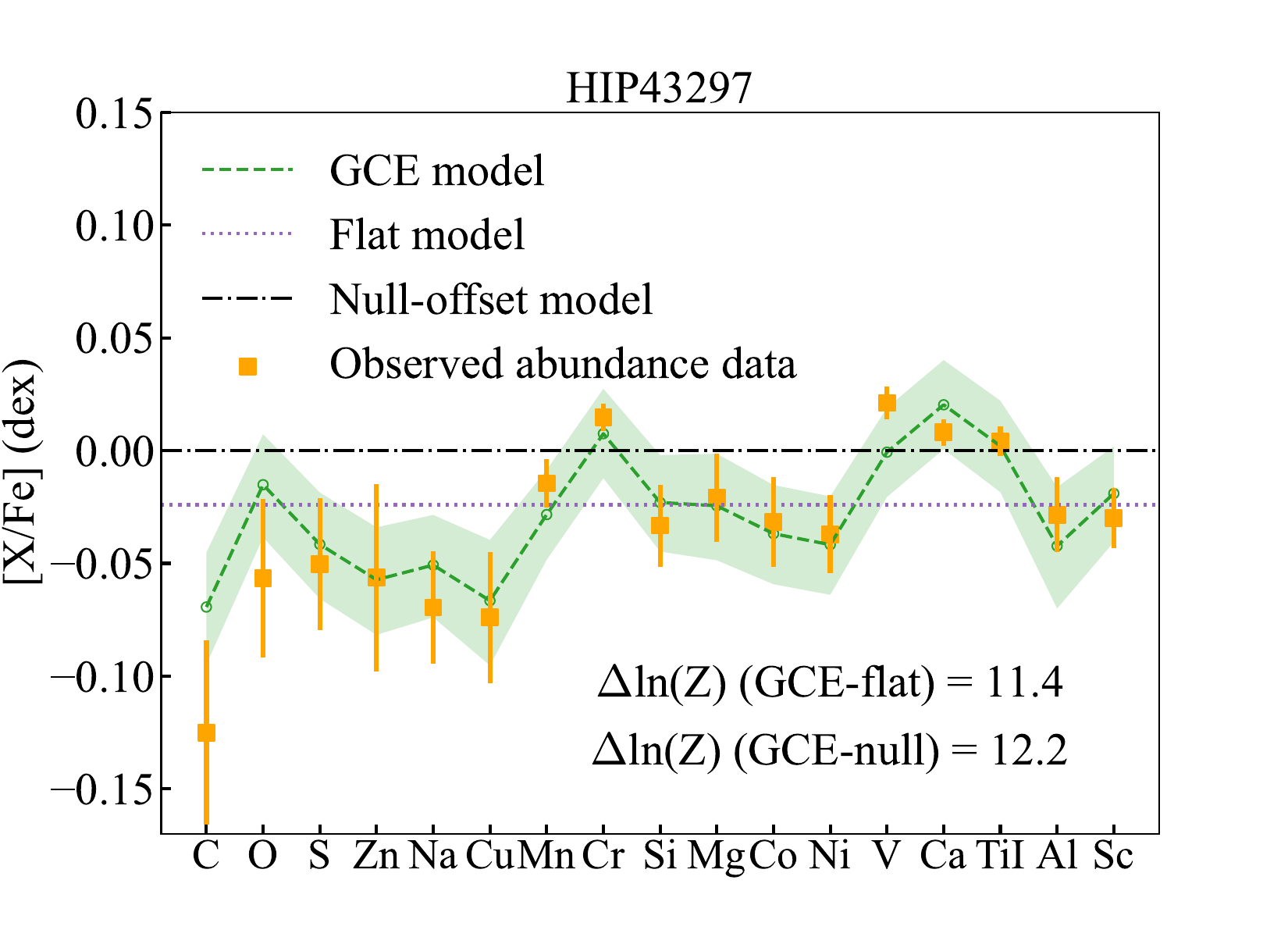}
    \caption{GCE-uncorrected elemental abundances [X/Fe] for HIP\,43297 which is well described by the GCE model. The observed abundance data are shown in orange, while the best-fitting abundance patterns for the flat model, null-offset, and GCE model are shown by the dotted purple, dot-dashed black, and dashed green lines, respectively. The green shaded regions indicate the 2$\sigma$ posterior probability distributions of the GCE models.}
    \label{fig:bayesian_abundance_patterns_GCE}
\end{figure}

In order to understand the extent to which solar twin abundance patterns are shaped by GCE effects, we analyse the GCE-uncorrected [X/Fe] abundance data by examining the Bayesian evidence difference for the null-offset and flat models compared to the GCE model. An example model fit is illustrated in Figure \ref{fig:bayesian_abundance_patterns_GCE}, and the final numerical fitting results for the solar twin sample are presented in Table \ref{tab:gce_results}. 

\begin{table*}
\centering
\caption{Bayesian GCE and flat model results including the Bayesian evidence differences and best-fit parameters.}
\label{tab:gce_results}
\begin{tabular}{lcccc}
\hline
Star &
$\Delta \ln Z_{\rm (GCE-null)}$ &
$\Delta \ln Z_{\rm (GCE-flat)}$ &
GCE Model Age (Gyr) & $\delta_{\rm flat}$ (dex) \\
\hline
HIP10175  & 7.65  & 9.85  & $3.44^{+0.53}_{-0.59}$ & $-0.012^{+0.007}_{-0.008}$ \\
HIP101905 & 12.55 & 11.24 & $0.98^{+0.65}_{-0.59}$ & $-0.031^{+0.011}_{-0.013}$ \\
HIP102040 & 5.19  & 7.63  & $3.49^{+0.74}_{-0.83}$ & $-0.012^{+0.009}_{-0.011}$ \\
HIP102152 & 1.61  & 5.88  & $6.14^{+0.39}_{-0.37}$ & $0.002^{+0.003}_{-0.003}$ \\
HIP10303 & -5.14 & -2.15 & $5.98^{+0.90}_{-0.83}$ & $0.007^{+0.005}_{-0.006}$ \\
\vdots       & \vdots & \vdots & \vdots & \vdots \\
\hline
\end{tabular}
\begin{flushleft}
\textit{Note.} The uncertainties represent the 16th and 84th percentile bounds 
of the posterior distribution. The full version of this table is available in electronic form.
\end{flushleft}
\end{table*}

Guided by the interpretive framework of \cite{kass_raftery_1995}, we adopt the following criteria to identify stars with surface compositions positively supported by GCE effects, chosen such that fewer than 0.7\% of mock noise realisations exceed both thresholds (corresponding to the 99.3rd percentile of each mock noise distribution); 
\begin{enumerate}[label=(\roman*)]
    \item $\Delta \ln Z_{\rm (GCE-null)}$ $>$ 2;
    \item $\Delta \ln Z_{\rm (GCE-flat)}$ $>$ 4.
\end{enumerate}

\begin{figure}
    \centering
    \includegraphics[width=1\linewidth]{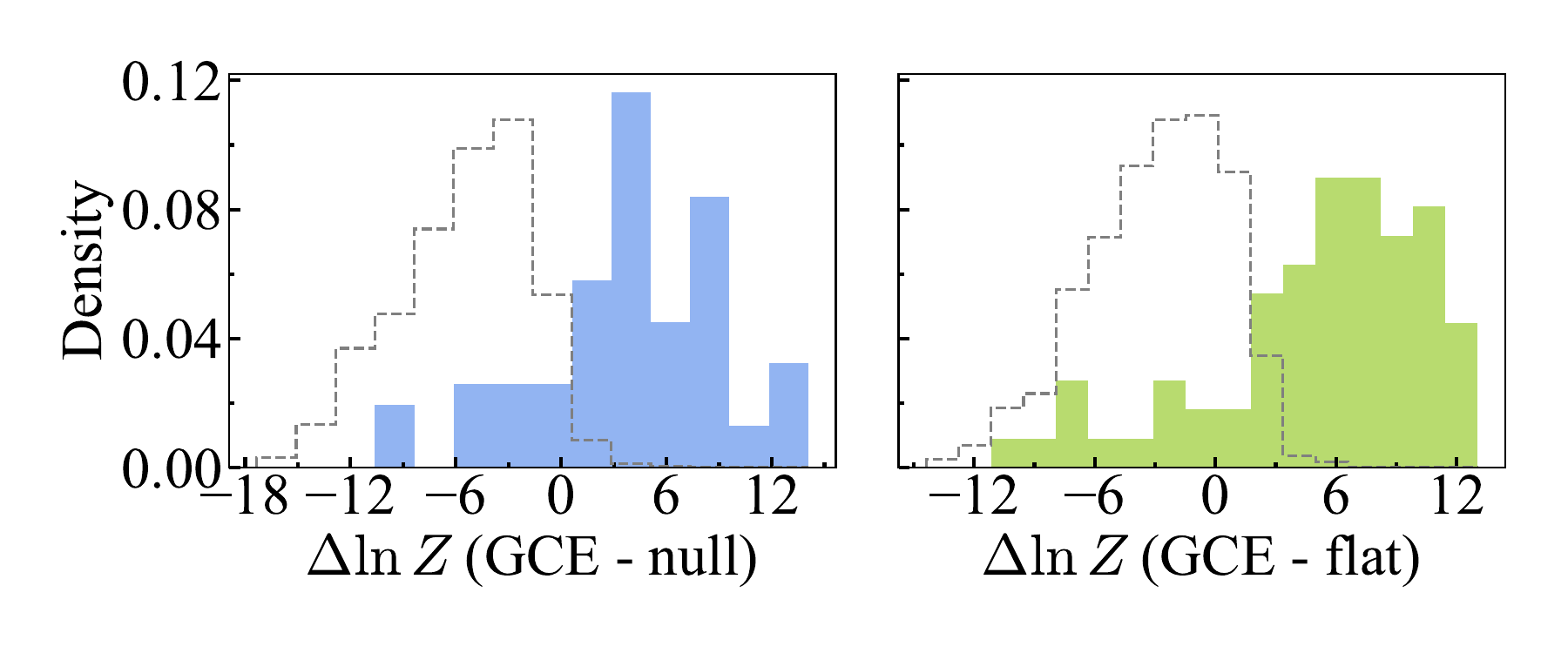}
    \caption{Distributions of the difference in Bayesian evidence $\Delta \ln Z$ between the GCE model and the null-offset (left) and flat models (right) fitted to the observed [X/Fe] GCE-uncorrected data for the sample of 69 solar twins in the solid colour. The dashed grey lines represents the equivalent distributions for model fits to corresponding mock noise samples.}
    \label{fig:bayesian_evidence_difference_GCE}
\end{figure}

The probability of obtaining an abundance pattern with Bayes factor meeting the above criteria under the baseline models is therefore $p \leq 0.007$.

Across our sample, 48 solar twins satisfy criterion (i), 47 satisfy criterion (ii), and 43 satisfy both of the above criteria. Applying binomial statistics to this result yields a probability of $\sim10^{-74}$ of observing at least 43 such events by chance. This demonstrates that the large number of stars consistent with our Bayesian GCE model is unlikely to occur randomly, supporting the interpretation that the inferred trends are physical in origin. Treating the 43/69 stars as a binomial fraction, we find that approximately 62.3\,$\pm$\,5.8\% of the solar twins in our sample have abundance patterns well-described by Galactic chemical evolution effects.

\subsubsection{Planet Engulfment Signatures}\label{subsubsec:planet_engulfment_signatures}
\begin{figure}
    \centering
    \includegraphics[width=1\linewidth]{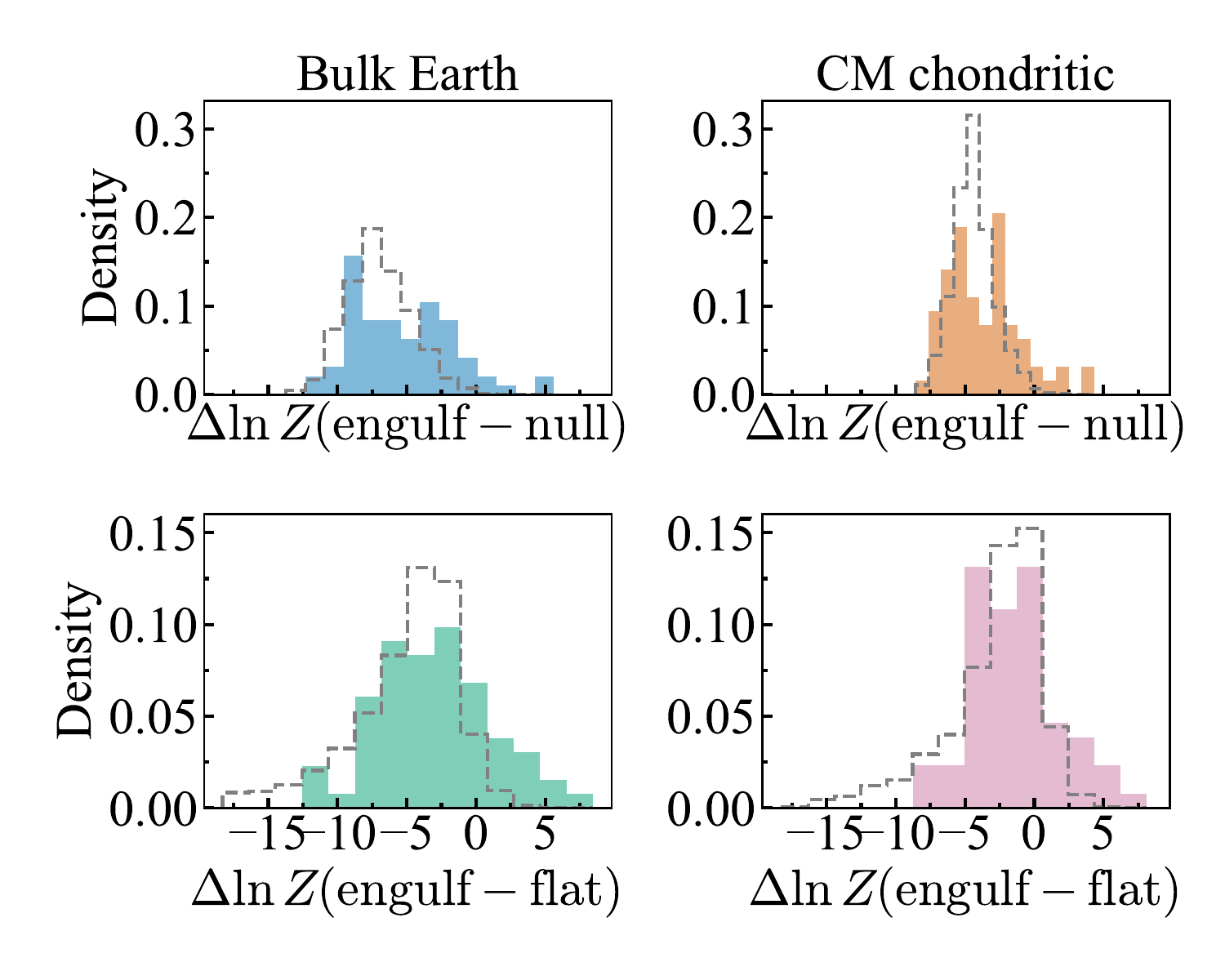}
    \caption{Distributions of the difference in Bayesian evidence $\Delta ln(Z)$ between the planetary engulfment models and the null-offset and flat models fitted to the observed [X/Fe] GCE-corrected (B18) data for the sample of 69 solar twins in the solid colour. The dashed grey lines represents the equivalent distributions for model fits to corresponding mock noise samples.}
    \label{fig:bayesian_evidence}
\end{figure}

Planetary engulfment candidates were identified by comparing the Bayesian evidence for the null-offset and flat models against that of the bulk Earth and CM chondritic engulfment models, using the GCE-corrected (B18) observed [X/Fe] abundances and the corresponding mock noise datasets. The resulting $\Delta \ln Z$ distributions are shown in Figure \ref{fig:bayesian_evidence}, with final Bayesian fitting results presented in Table \ref{tab:engulf_results}. 

\begin{table*}
\centering
\caption{Bayesian engulfment model results, including the Bayesian evidence differences and best-fit parameters, for four engulfment candidates identified in Section \ref{subsubsec:planet_engulfment_signatures}. }
\label{tab:engulf_results}
\begin{tabular}{lcccccc}
\hline
Star &
$\Delta \ln Z_{\rm (CM-null)}$ &
$\Delta \ln Z_{\rm (CM-flat)}$ &
$M_{\rm CM}\,(M_\oplus)$ &
$\Delta \ln Z_{\rm (BE-null)}$ &
$\Delta \ln Z_{\rm (BE-flat)}$ &
$M_{\rm BE}\,(M_\oplus)$ \\
\hline
HIP101905 & 4.33 & 8.10 & $7.39^{+2.46}_{-2.26}$ & 4.69 & 8.45 & $4.65^{+1.42}_{-1.24}$ \\

HIP30502  & 3.64 & 3.65 & $4.35^{+1.46}_{-1.33}$ & 5.63 & 5.65 & $2.79^{+0.73}_{-0.63}$ \\

HIP77052 & 2.06 & 5.34 & $5.35^{+2.25}_{-2.13}$ & 1.46 & 4.74 & $2.56^{+0.92}_{-0.80}$ \\

HIP85042 & 1.87 & 4.86 & $3.45^{+1.36}_{-1.14}$ &
1.54 & 4.53 & $1.91^{+0.62}_{-0.53}$ \\

\vdots & \vdots & \vdots &
\vdots & \vdots & \vdots & \vdots \\
\hline
\end{tabular}
\begin{flushleft}
\textit{Note.} The uncertainties represent the 16th and 84th percentile bounds 
of the posterior distribution. The full table, including the complete solar twin sample, is available in electronic form.
\end{flushleft}
\end{table*}

Similar to our analysis of GCE effects, we defined criteria for evidence of planetary engulfment based on the interpretive framework of \cite{kass_raftery_1995} and chosen to correspond to the 99.6th percentile of the mock noise $\Delta \ln Z$ distributions (Figure~\ref{fig:bayesian_evidence}). We note that our thresholds are intentionally more permissive than the strong evidence criteria of \cite{kass_raftery_1995}, as our goal is to identify candidates for further follow-up. For both the CM chondritic and bulk Earth engulfment models, we require:
\begin{enumerate}[label=(\roman*)]
    \item $\Delta \ln Z_{\rm (engulf-null)}$ $>$ 1;
    \item $\Delta \ln Z_{\rm (engulf-flat)}$ $>$ 3.6.
\end{enumerate}

Both the CM chondritic and bulk Earth engulfment criteria recover the same four solar twins: HIP\,101905, HIP\,30502, HIP\,77052, and HIP\,85042. Assuming binomial statistics and adopting the corresponding mock-noise p-values (see Appendix \ref{apdx:pvalues}), the probability of recovering at least four stars satisfying the engulfment criteria under the baseline models is $\sim10^{-7}$ for the bulk Earth model and $\sim10^{-4}$ for the CM chondritic model. This indicates that recovering all four engulfment candidates by chance is unlikely, supporting the interpretation that the inferred abundance trends are physical in origin. For all four stars, the observed abundance patterns are at least as significant as the most extreme realisation in the mock-noise sample under the null-offset model, corresponding to $p \leq 0.001$, while the equivalent probabilities under the flat model are $p \leq 0.004$.

\begin{figure}
    \centering

    \begin{subfigure}{\linewidth}
        \centering
        \includegraphics[width=\linewidth]{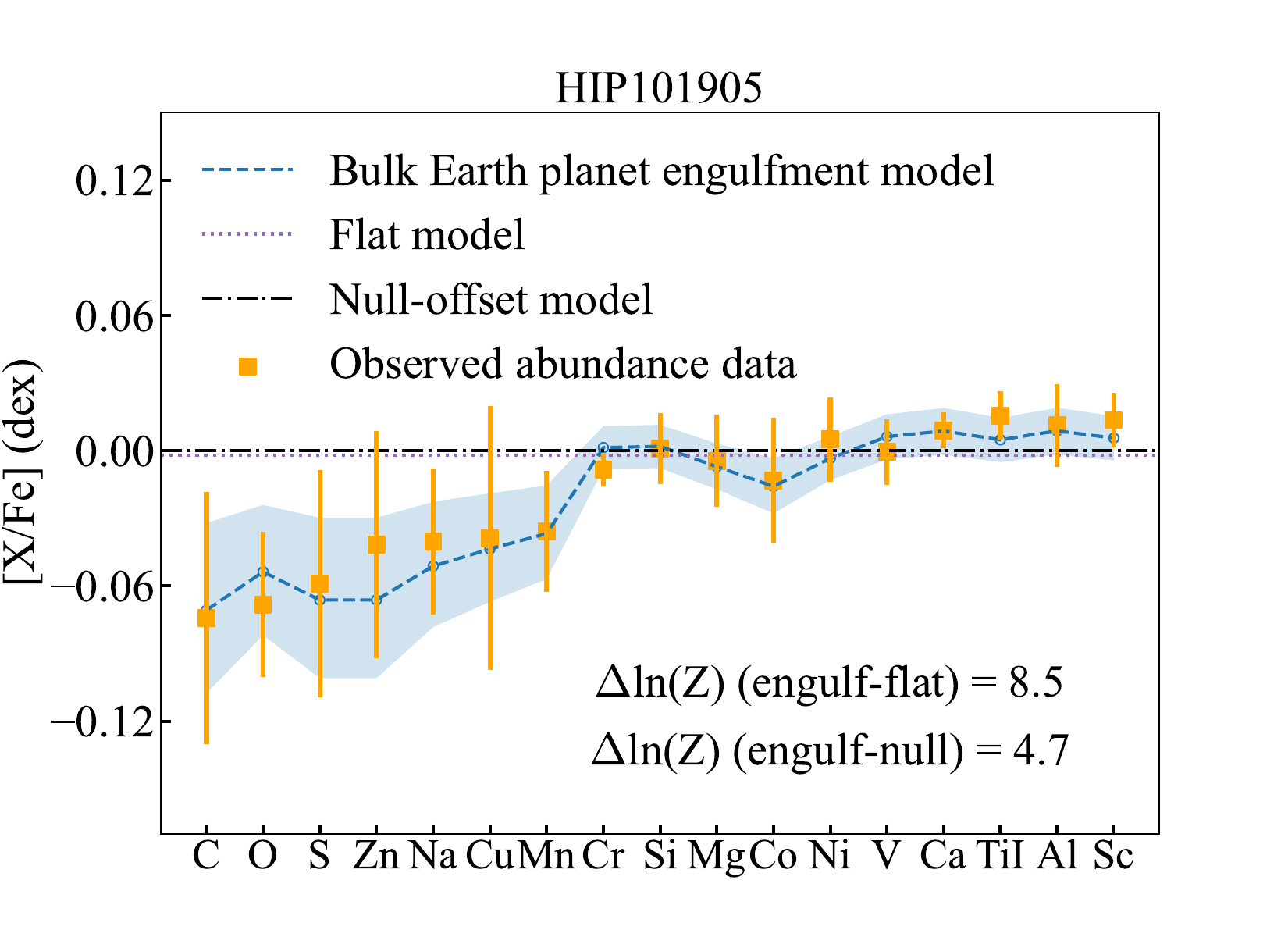}
    \end{subfigure}

    \begin{subfigure}{\linewidth}
        \centering
        \includegraphics[width=\linewidth]{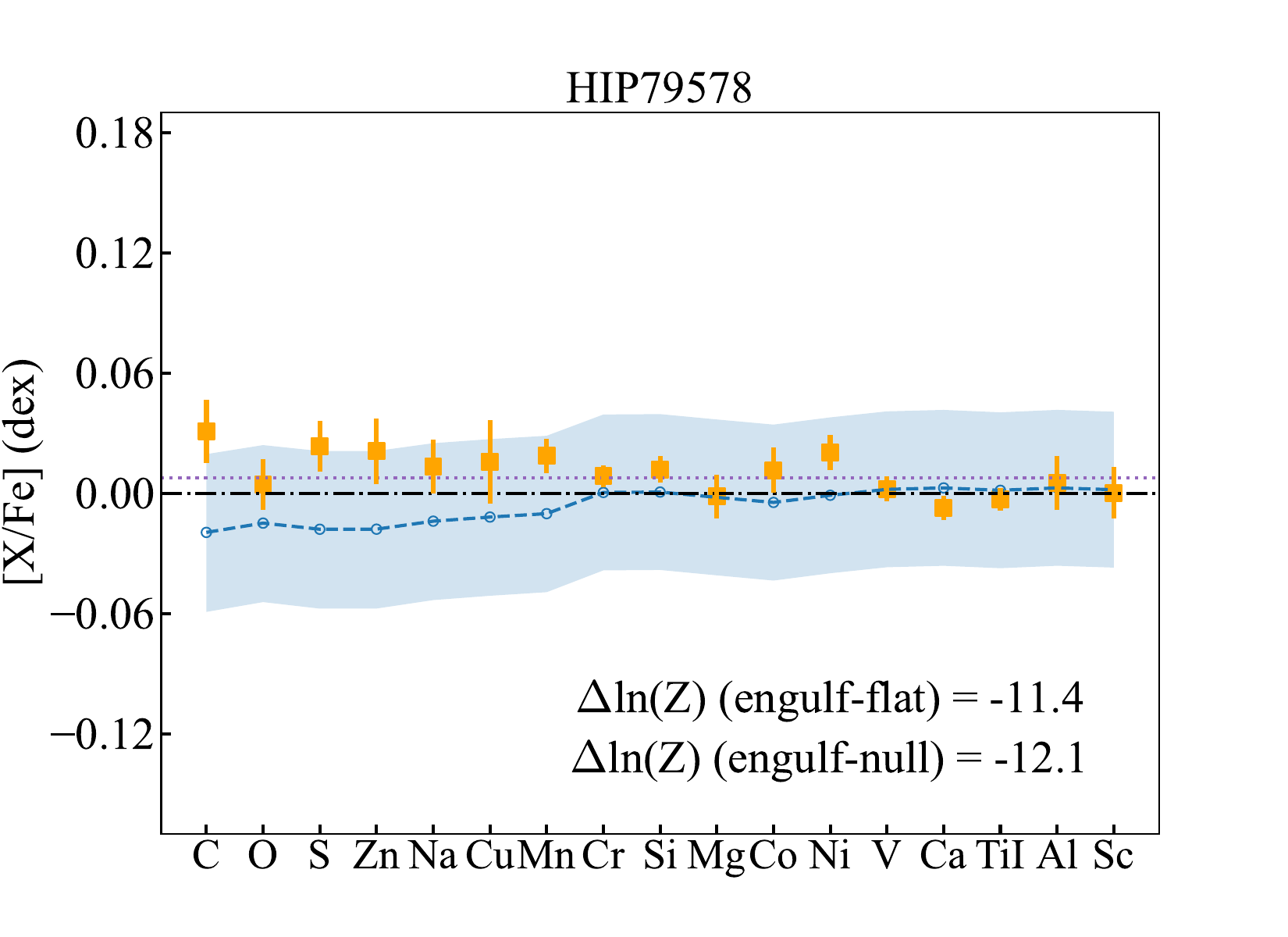}
    \end{subfigure}

    \caption{Bayesian fitting results for the GCE-corrected (B18) elemental abundances [X/Fe] of two solar twins: the best engulfment candidate HIP\,101905 (top) and a representative non-candidate, HIP\,79578 (bottom), comparing the baseline null-offset and flat models to the Bulk Earth engulfment model. The observed abundance data are shown in orange, while the best-fitting abundance patterns for the flat model, null-offset, and engulfment models are shown by the dotted purple, dot-dashed black, and dashed lines, respectively. The shaded regions indicate the 2$\sigma$ posterior probability distributions of the engulfment models.}

    \label{fig:bayesian_engulfment_abundance_patterns}
\end{figure}

The star exhibiting the strongest evidence for engulfment across all models is HIP\,101905. Under the bulk Earth engulfment model, this solar twin is inferred to have accreted $4.65^{+1.42}_{-1.24}\, M_{\oplus}$ of bulk Earth material, alongside the minimum allowed contribution of 0.1\,$M_{\oplus}$ of CM chondritic material. Conversely, under the CM chondritic engulfment model, it is inferred to have accreted $7.39^{+2.46}_{-2.26}\,M_{\oplus}$ of CM chondritic material, together with the minimum allowed 0.1\,$M_{\oplus}$ of bulk Earth material. Fitting results for HIP\,101905 under the bulk Earth engulfment model 
are shown in Figure~\ref{fig:bayesian_engulfment_abundance_patterns}, alongside a star without evidence of planetary ingestion for reference. The corresponding HIP\,101905 CM chondritic fit is presented in Appendix~\ref{apdx:bayesian_engulfment_fitting}. The inferred accreted masses for the remaining three engulfment candidates are listed in Table \ref{tab:engulf_results}, with 
corresponding Bayesian fitting plots also presented in Appendix~\ref{apdx:bayesian_engulfment_fitting}.

To assess the robustness of our candidate identifications to the choice of GCE correction, we repeated the analysis using the GCE-corrected (TW) abundances (see Appendix \ref{apdx:bayesian_engulfment_myGCEcorrection} for detailed results and discussion). Using the same selection criteria outlined above, HIP\,101905 and HIP\,77052 are recovered as candidates under both the GCE (TW) and (B18) corrections. HIP\,30502 and HIP\,85042 are not recovered with the GCE-corrected (TW) data, suggesting their identification is sensitive to the adopted GCE trends. Two additional candidates HIP\,22263 and HIP\,40133 emerge under our GCE-correction (TW). Based on these results, we identify HIP\,101905 and HIP\,77052 as highly probable engulfment candidates, while HIP\,85042, HIP\,30502, HIP\,22263 and HIP\,40133 are classified as tentative engulfment candidates.
\section{Discussion}\label{sec:discussion}
\subsection{Implications of Bayesian Results}\label{subsec:implications_bayesian}
Our Bayesian analysis (Sections~\ref{sec:bayesian_method} and \ref{sec:GCE_bayesian_results}) indicates that the majority (62.3\,$\pm$\,5.8\%) of solar twins in our sample exhibit abundance patterns well described by GCE effects, while the remainder are likely too chemically similar to the Sun for GCE signatures to be robustly disentangled from intrinsic abundance scatter. Extending this analysis to GCE-corrected abundance data, we find that only 2--6 stars in our sample exhibit abundance patterns that significantly deviate from the flat and null-offset baseline models (Section~\ref{subsubsec:planet_engulfment_signatures}). The remaining solar twins show abundance patterns insufficiently distinct to be separated from baseline behaviour. Together, these results suggest that solar twins possess compositions that are preferentially explained by GCE trends and intrinsic abundance differences over planet ingestion, hence implying that the Sun may not be chemically peculiar after accounting for these effects.

Our findings sit within an ongoing debate over whether the Sun's apparent chemical peculiarity can be explained by GCE effects, with different modelling frameworks achieving different results. Studies employing linear, empirically-derived abundance correlations (e.g., \citealp{nissen_2015, nissen_2016, spina_planet_2016, bedell_2018, martos_2025, ghezzi_2026}) found that the solar $T_{\mathrm{cond}}$ trend persists even after accounting for GCE effects, suggesting that the Sun remains chemically anomalous relative to $\sim$70--90\% of solar twins. However, more recent work has challenged these findings. For example, \cite{rampalli_galactic_2026} used theoretical models of core-collapse and Type Ia supernova enrichment to show that the Sun is chemically ordinary, with abundances consistent with the intrinsic scatter expected from GCE trends. Similarly, \cite{cowley_galactic_2022} found that the refractory depletion trend is sensitive to the statistical methods adopted for the GCE correction.

We additionally note that definitions of the solar chemical peculiarity in previous studies have largely relied on trends with $T_{\mathrm{cond}}$, an approach inherently limited by uncertainties in condensation temperatures. Studies such as \cite{lodders_2003, spaargaren_2025} have shown that elemental condensation temperatures are not universal, but depend sensitively on the chemical composition of the protoplanetary disk. In particular, the abundances of carbon and oxygen influence the volatility of planet-forming elements such as Fe, Mg, Si, Ca, Al, Na, Ni, and S. For example, in oxygen-poor conditions, oxygen becomes preferentially locked in CO gas, suppressing silicate and oxide formation and thereby altering the condensation sequence. These results suggest that the commonly adopted Solar System condensation temperature scale \citep{lodders_2003} may not be universally applicable to exoplanetary systems, and that interpretations of linear stellar abundance trends based on fixed $T_{\mathrm{cond}}$ values should be treated with caution. Motivated by these limitations, our work instead examines solar twin chemistry independently of $T_{\mathrm{cond}}$ values.

Although previous studies and the results presented here have not yet converged on a unified statement regarding the Sun's anomalous composition, together they emphasise the sensitivity of this analysis to the various methods used to characterise solar twin abundance trends and GCE effects. These findings highlight that robust treatment of GCE effects is essential when analysing solar twin abundances, and represents a necessary step towards reliably isolating the signatures of other processes that shape stellar composition.

As discussed in Section~\ref{subsubsec:planet_engulfment_signatures}, we identify 2--6 potential engulfment candidates from the 69 stars analysed for planetary ingestion signatures within our Bayesian framework. This corresponds to an inferred planetary ingestion occurrence rate of $\sim$3--9\%, ($\sim$1--13\% accounting for Poisson noise), comparable to rates derived in recent observational studies and theoretical predictions for Sun-like stars. Theoretical work by \cite{behmard_2023a, oconnor_lai_2025} similarly predicts detectable engulfment occurrence rates of $\sim$2--5\% in single Sun-like stars, based on detailed stellar evolution modelling, analytic descriptions of ultra-short-period planet evolution, and models of violent dynamical interactions. Similarly, observational studies of multi-star systems by \cite{liu_2024, behmard_2023a} report engulfment occurrence rates in the range of $\sim$4--10\%. The consistency of our results, derived from a sample of field solar twins, with these observational estimates may suggest that rates of dynamical instability in non-co-natal Sun-like stars are broadly similar to those in co-natal, twin binary systems.

Our engulfment candidates span a broad age range according to age estimates by \cite{spina_2018}, which has implications for the timing of engulfment events. While planetary ingestion can occur any time during a star’s evolution, it is expected to be most common within the first hundred Myr of stellar system formation due to enhanced dynamical activity and instability \citep{pinsonneault_2001, bitsch_2023, huhn_bitsch_2023}. Detailed stellar evolution models presented by \cite{behmard_2023a, behmard_planet_2023} further show that observable chemical signatures of engulfment diminish after the ingestion event as a result of internal mixing processes, typically falling below detectable levels within 2\,Gyr for early (near-ZAMS) engulfment in solar-like stars. Later-stage engulfment can produce long-lived signatures of up to $\sim$1.5\,Gyr. Our most promising candidate, HIP\,101905, and tentative candidate HIP\,22263, have ages of 1.2\,Gyr and 0.8\,Gyr respectively, consistent with enrichment from early engulfment. The remaining candidates are older ($\gtrsim$4.5\,Gyr), making such signatures less likely under both dynamical and stellar evolution models. Nevertheless, such events remain plausible given that engulfment may occur relatively late in the main-sequence phase, and also taking into consideration the large uncertainties in stellar age estimates (1--2\,Gyr). Although our candidate sample is not large enough for robust conclusions, we emphasise that the evolution of engulfment signatures with stellar age is an important avenue for future observational study.

\subsection{Limitations \& Future Work}\label{subsec:limitations_futurework}
In this section we will discuss several limitations of this work. Firstly, we have not taken into account the influence of magnetic activity on solar twin photospheres. High rates of chromospheric activity, typically found in young stars ($\lesssim$1–5\,Gyr), can alter atmospheric structure and broaden spectral features, potentially leading to overestimated microturbulence and underestimated effective temperature and chemical abundances \citep{spina_2020, cao_star-spots_2022}. Recent studies suggest that star spots can introduce scatter in abundance measurements on the order of 0.05\,dex, comparable to the effects of atomic diffusion, and can be partially entangled with GCE trends (e.g., \citealp{spina_2020, wilson_stellar_2023}). Moreover, there is tentative evidence that chromospheric activity alone can induce $T_{\rm cond}$ trends in abundance patterns that mimic planet-related signatures \citep{yu_c3po_2025}.

\begin{figure}
    \centering
    \includegraphics[width=0.7\linewidth]{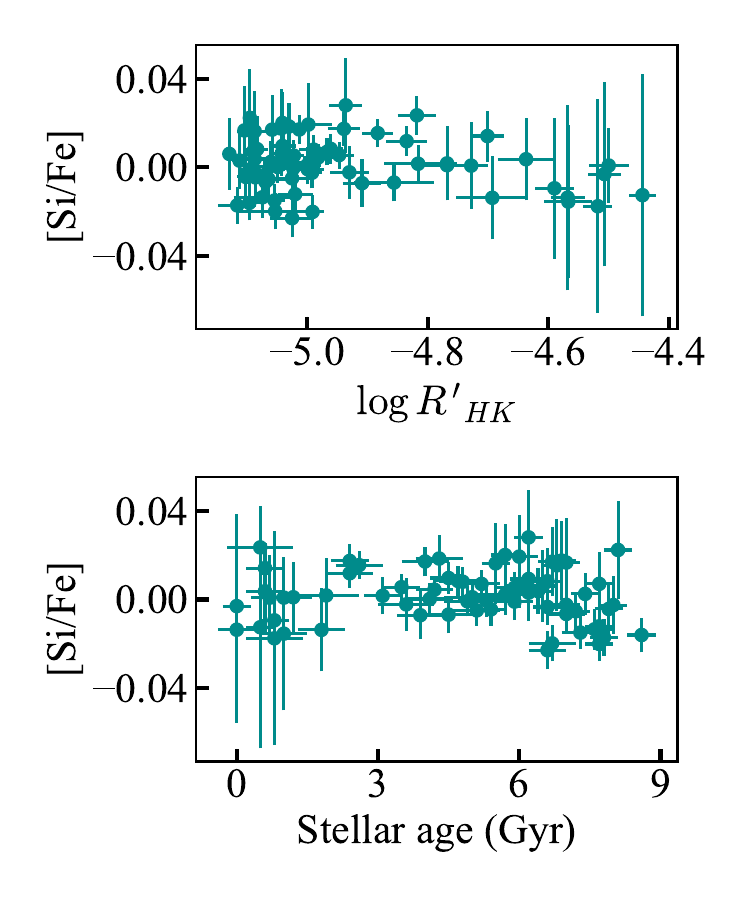}
    \caption{[Si/Fe] GCE-corrected (B18) abundances of representative species silicon as a function of magnetic activity index $logR'_{HK}$ (top) and stellar age (bottom). We observe no strong correlations between these variables, with similar results found for other elemental abundances.}
    \label{fig:silicon_magnetic}
\end{figure}

\begin{figure}
    \centering
    \includegraphics[width=0.7\linewidth]{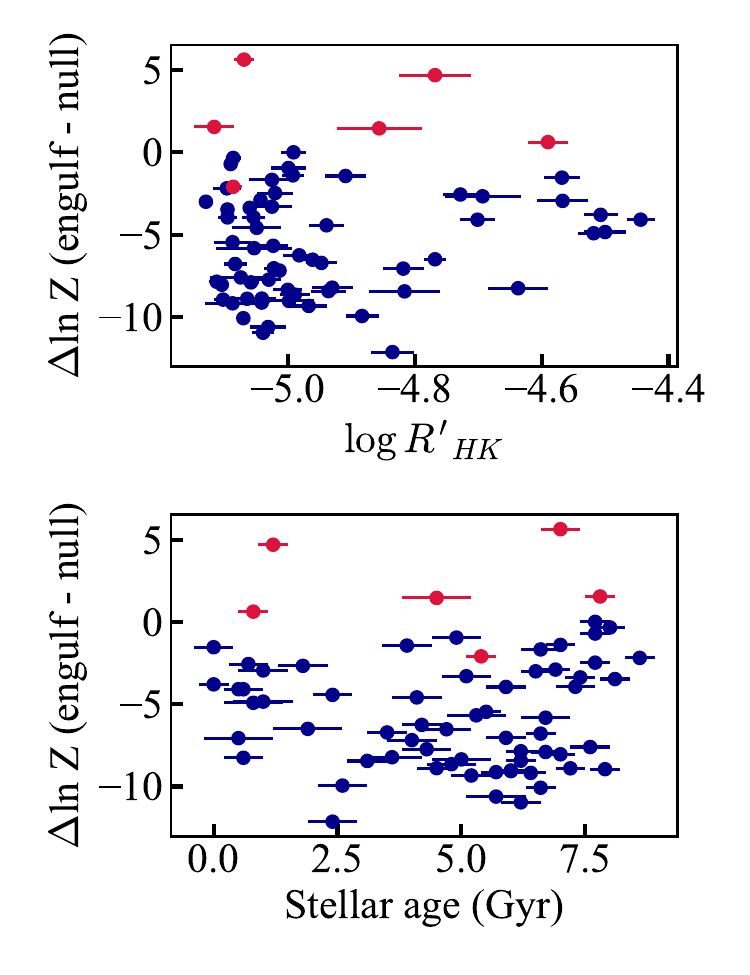}
    \caption{Bayesian evidence differences comparing the bulk Earth engulfment model to the null-offset model, as a function of magnetic activity index $logR'_{HK}$ (top) and stellar age (bottom). No strong correlations are observed between these quantities, and a consistent result is found when comparing to the flat model and the CM chondritic engulfment model. The six engulfment candidates are shown in red, while the remaining stars are shown in dark blue.}
    \label{fig:delta_lnZ_magnetic_activity}
\end{figure}

To investigate the influence of magnetic activity on our solar twin sample, we employ the time-averaged magnetic activity indices, $logR'_{HK}$, derived for the stars used in this study by \cite{carvalho-silva_new_2025}. We examine correlations between magnetic activity, stellar age, elemental abundances, and Bayesian evidence difference, with the results presented in Figures \ref{fig:silicon_magnetic} and \ref{fig:delta_lnZ_magnetic_activity}. We find that abundance uncertainties increase toward younger ages and higher levels of activity, which is expected, as enhanced magnetic activity leads to increased line broadening and consequently greater scatter in line-by-line differential abundance measurements. Beyond this trend, we find no strong correlations between magnetic activity and elemental abundances nor Bayesian evidence difference, and are therefore unable to quantitatively incorporate magnetic activity into our current analysis. Four of our six engulfment candidates have ages $\geq$4.5\,Gyr (Section \ref{subsec:implications_bayesian}) and are therefore expected to be relatively chromospherically quiet, with atmospheric parameters and abundance patterns only minimally affected by magnetic activity. While magnetic activity effects are beyond the scope of our current work, they are important to account for in future solar twin studies. 

Other limitations of this study arise from the simplifying assumptions adopted in our Bayesian analysis. We approximated the stellar convective zone mass as $M_{CZ} = f_{CZ}M_{star}=0.02\,M_{\odot}$ across the entire sample, however, this assumption is unlikely to be true for all stars, since $M_{CZ}$ depends on mass and age, which vary across our sample ($\sim$0.95--1.1\, $M_{\odot}$; ages $\sim$0--9\,Gyr). To therefore determine the extent to which $M_{CZ}$ may differ across our solar twin sample, we employ stellar evolution code MESA \citep{paxton_2011, jermyn_modules_2023} with default settings to evolve eight solar metallicity ($Z=0.02$) stars in the mass range 0.95--1.1\,$M_{\odot}$ through the main sequence, finding that $M_{CZ}$ varies over the range ${\sim}0.01$--$0.035\,M_{\odot}$. Applying these updated $M_{CZ}$ values to our engulfment models, we find that the inferred accreted mass varies by $\sim$1.7\,$M_{\oplus}$ on average, but may change by up to 6.6\,$M_{\oplus}$, as in the case of HIP\,40133. While the fixed convective zone mass represents a limitation of this work, it does not affect our primary conclusion regarding the engulfment occurrence rate, and impacts only the inferred accreted masses for stars whose properties differ noticeably from solar. Furthermore, uncertainties around stellar mass, age, and theoretical stellar modelling (e.g., \citealp{joyce_chaboyer_2018, cinquegrana_joyce_2022}) make it challenging to estimate $M_{CZ}$ accurately for individual stars. Therefore, our solar-value assumption remains a well-motivated default for a sample defined by its similarity to the Sun. Future analyses could marginalise over the value of $M_{CZ}$ as a free parameter in the models.

Our current engulfment model is additionally restricted to two compositions of planetary material and assumes instantaneous, homogeneous mixing of ingested material through the convective zone. Addressing these limitations provides promising directions for future work. This may include a case-by-case analysis of the two to six engulfment candidates identified here, using detailed stellar evolution modelling to better constrain the timing, mass, and composition of the accreted material responsible for the observed abundance signatures. Extending the models to incorporate a wider range of accreted compositions, as well as the internal mixing processes that may deplete surface abundances over time (see Section \ref{subsec:implications_bayesian}), would offer a deeper understanding of these candidates. More broadly, additional observational diagnostics could also be explored, such as searches for potential perturbers, to further test the engulfment scenario in these systems.

\section{Conclusion} \label{sec:conclusion}
In this study we re-analysed high-quality spectra for the sample of 79 solar twins presented by \cite{bedell_2018} and \cite{spina_2018} using the spectroscopic analysis tool \texttt{Korg}, different model atmospheres and independent statistical techniques. We performed a line-by-line differential analysis using equivalent-width techniques to derive atmospheric parameters and abundances for 18 elements, achieving an average abundance precision of 0.015\,dex (3.5\%). We find small systematic offsets between our results and those of \cite{bedell_2018}, typically $\lesssim$0.01\,dex, likely reflecting differences in the spectral analysis codes employed. Star-to-star scatter in the abundance differences is smaller than the combined measurement uncertainties for all 17 species, confirming close agreement with \cite{bedell_2018} at the individual-star level. 

A Bayesian indicator was implemented to characterise the impact of Galactic chemical evolution on the composition of the 69 non-$\alpha$-enhanced stars, and we found that accounting for GCE effects is necessary to describe 62.3$\pm$5.8\% of our sample. Applying our Bayesian framework to the GCE-corrected [X/Fe] 
abundances, we disentangled the contributions of planetary engulfment, intrinsic abundance scatter, and overall abundance offsets to the observed stellar abundance patterns. We identify two stars (HIP\,101905 and HIP\,77052) that show good evidence for planetary engulfment while four (HIP\,85042, HIP\,30502, HIP\,40133 and HIP\,22263) show tentative evidence dependent on the GCE correction applied. This corresponds to an engulfment occurrence rate of 3--9\% (1--13\% accounting for Poisson uncertainties). These systems represent promising targets for future detailed investigation. Overall, our results highlight the importance of accounting for GCE effects when interpreting solar twin abundance patterns, and suggest that, once these effects are properly mitigated, the Sun may not be chemically peculiar relative to the average solar twin.

\begin{acknowledgments}
We are grateful to the anonymous referee for their comments which have helped us improve this paper. We thank Andrew R. Casey, Riley Thai and Shun Y. Cheung for valuable discussions and input. MB and LP are supported by the Commonwealth through an Australian Government Research Training Program Scholarship [DOI: https://doi.org/10.82133/C42F-K220]. FL was supported by the National Natural Science Foundation of China. MB, IM and LP acknowledge support from the Australian Research Council (ARC) Centre of Excellence for Gravitational Wave Discovery (OzGrav) through project number CE230100016. SXW and ZC acknowledge support from MOST grant 2025YFE0102100. The Flatiron Institute is a division of the Simons Foundation. 
\end{acknowledgments}

%

\software{\texttt{numpy} \citep{numpy}, \texttt{matplotlib} \citep{matplotlib}, \texttt{Korg} \citep{korg, korg2}.}



\bibliography{PhD_bib}{}
\bibliographystyle{aasjournalv7}



\appendix

\section{Mock noise samples}\label{apdx:mocknoise_sample}
\begin{figure}
    \centering
    \includegraphics[width=0.8\linewidth]{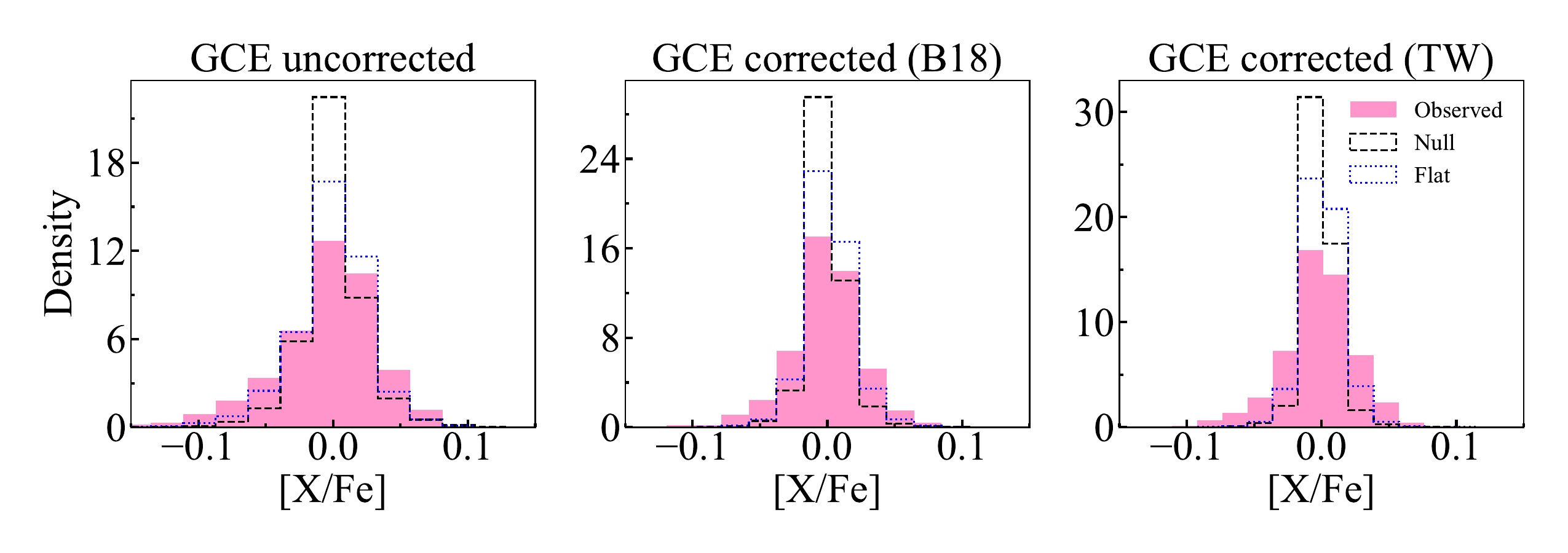}
    \caption{Comparison of the observed elemental abundances [X/Fe] (solid pink) with mock noise samples generated from the null-offset (dashed black line) and flat (dotted blue line) models. The comparison is shown for the GCE-uncorrected sample (left), and the two GCE-corrected samples using the \cite{bedell_2018} trends (middle) and our own age-abundance trends (right). Each distribution includes all measured species as well as all stars \& samples.}
    \label{fig:mocknoise_samples}
\end{figure}

In Figure \ref{fig:mocknoise_samples}, we compare the distribution of observed elemental abundances [X/Fe] for all 69 non-$\alpha$-enhanced stars with the elemental abundances [X/Fe] of the mock noise samples generated to represent the null-offset and flat models. Although the mock samples exhibit narrower distributions, they broadly capture the overall shape and range of the observed abundance distribution, making them suitable approximations of the baseline model behaviour. These results are consistent across the GCE-corrected and uncorrected data.

\section{GCE age-abundance trends}\label{apdx:GCEageabundancetrends}
Following \cite{bedell_2018}, we fit linear age-abundance trends to the [X/Fe] abundances derived in this work for the 69 non-$\alpha$-enhanced stars, using the stellar ages from \cite{spina_2018}. For each species, we maximise the log-likelihood function incorporating measurement uncertainties on both age and abundance, with three free parameters: age-abundance slope \textit{m}, intercept \textit{b}, and intrinsic scatter \textit{s}:

\begin{equation}
    \log \mathcal{L} (m,b, s) = -\frac{1}{2}\sum^N_{i=1} \frac{[y_i+mx_i-b]^2}{\sigma_{yi}^2+m^2\sigma_{xi}^2+s^2} -\frac{1}{2}\sum^N_{i=1}\log (\sigma_{yi}^2+m^2\sigma_{xi}^2+s^2)+\rm const
\end{equation}

where $x_i$ and $y_i$ are the age and [X/Fe] abundance of the
i-th star, $\sigma_{x_i}$ and  $\sigma_{y_i}$ are their respective measurement uncertainties, and the sum runs over all
N stars.

\begin{figure}
    \centering
    \includegraphics[width=\linewidth]{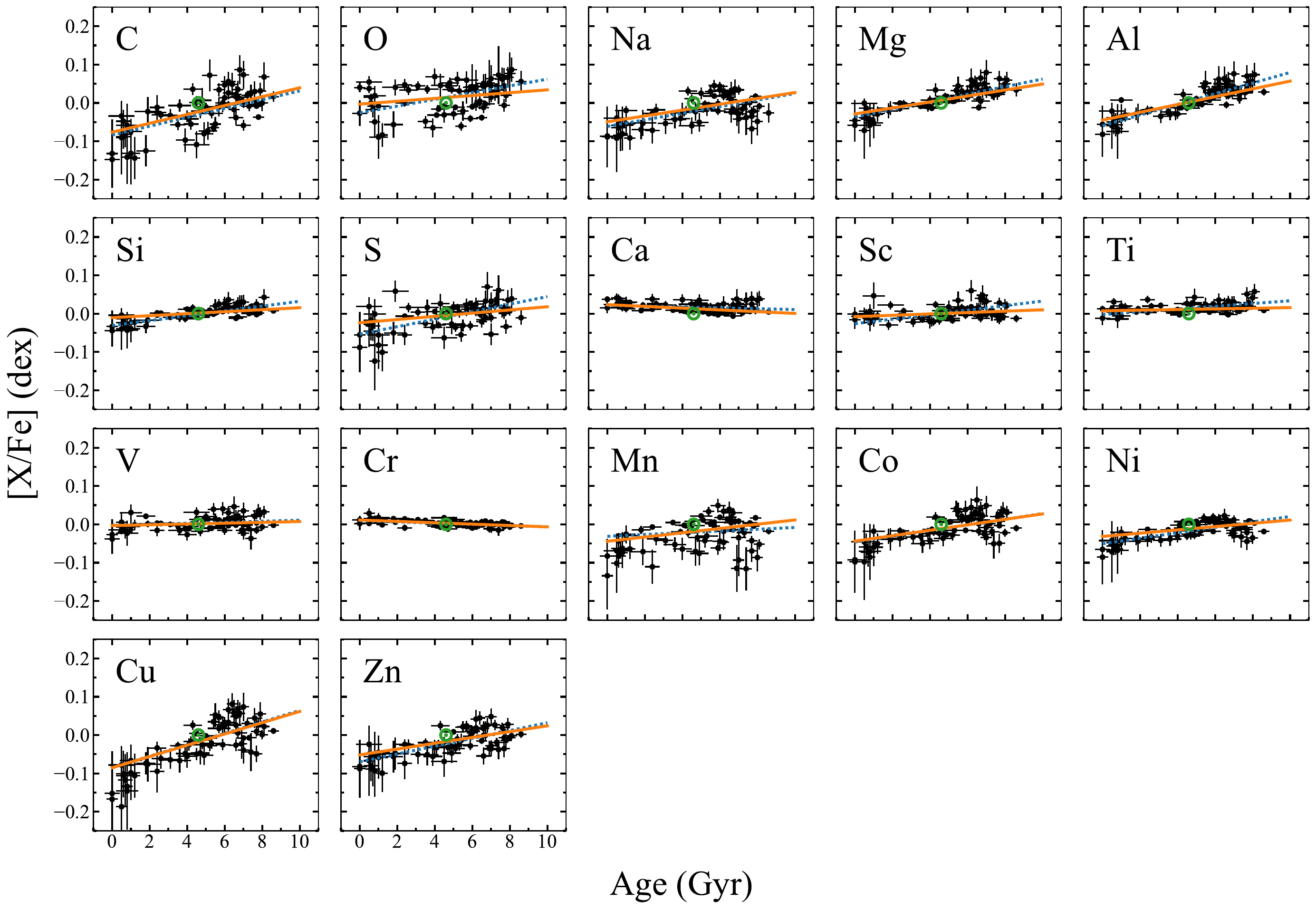}
    \caption{Elemental [X/Fe] abundances derived in this work as a function of stellar age. The solid orange lines show the best-fit linear age--abundance relations derived in this work, and the dotted blue lines show the corresponding relations from \cite{bedell_2018} for comparison. The solar abundances (green) were not considered in the fit, but are consistent with it.}
    \label{fig:myGCEabundanceagetrends}
\end{figure}

\begin{figure}
    \centering
    \includegraphics[width=0.5\linewidth]{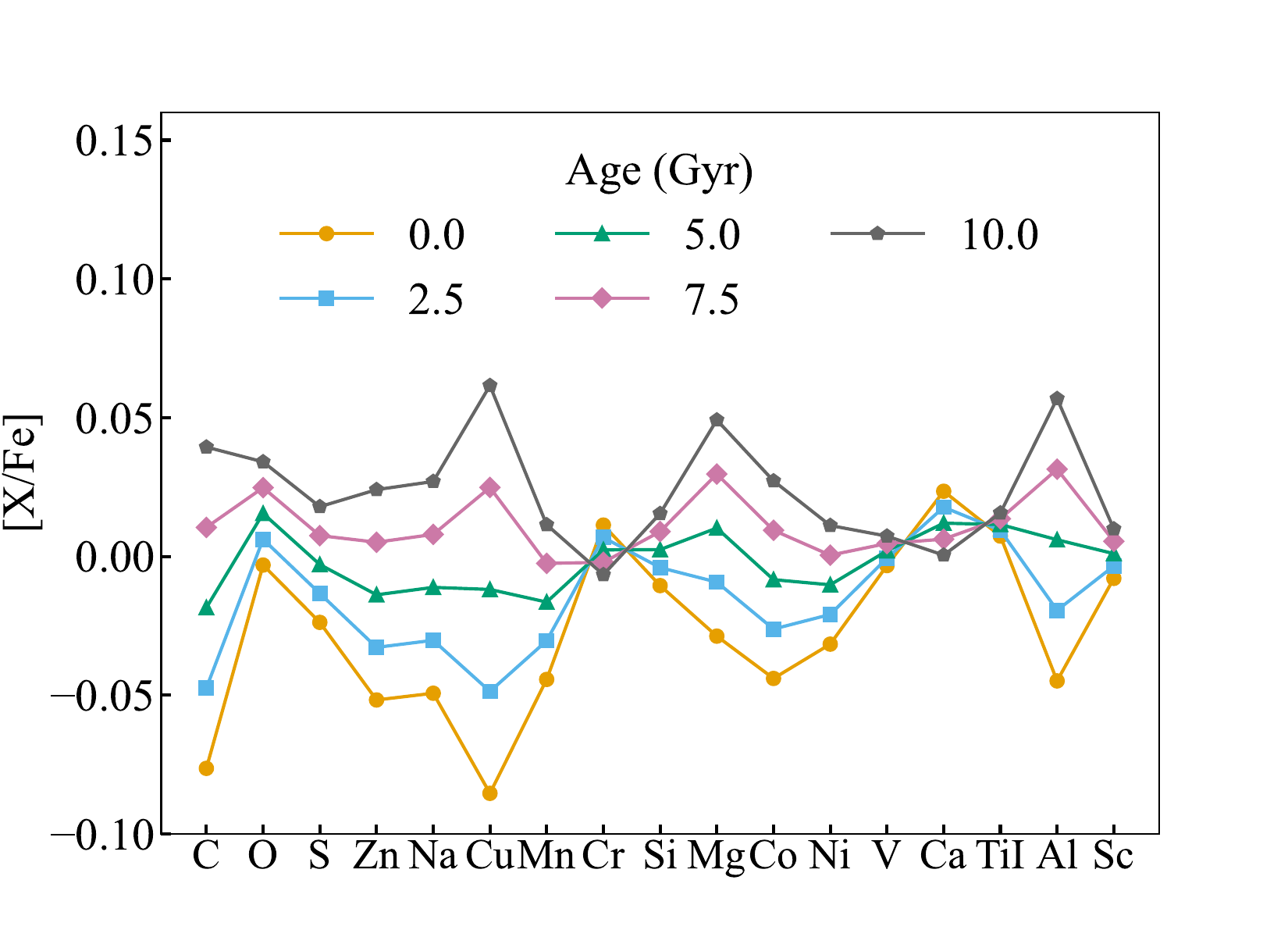}
    \caption{Predicted [X/Fe] abundance patterns for solar twins of different ages, based on the GCE age--abundance relations derived in this work. Abundance patterns for stars at ages 0, 2.5, 5, 7.5 and 10\,Gyr have been stylised with yellow circles, blue squares, green triangles, pink diamonds, and grey pentagons respectively.}
    \label{fig:myGCEabundanceagetrends_abundancepatterns}
\end{figure}

Best-fit parameters are obtained using the L-BFGS-B algorithm implemented in \texttt{scipy.optimize.minimize}. Parameter uncertainties are estimated from a grid search over the likelihood surface, defined as the mean of the 90\% confidence interval bounds. The resulting best-fit GCE trends are shown in Figure \ref{fig:myGCEabundanceagetrends} and the best-fit parameters with corresponding uncertainties are given in Table \ref{tab:myGCEabundanceagetrends}. Examples of the predicted abundance patterns for stars at different ages, based on the GCE trends derived in this work, are illustrated 
in Figure~\ref{fig:myGCEabundanceagetrends_abundancepatterns}.

\begin{table}
\centering
\caption{Best-fit parameters for GCE age-abundance trends based on the abundances derived in this work.}
\label{tab:gce_fits}
\begin{tabular}{lccc}
\hline\hline
Species & $m$ (dex Gyr$^{-1}$) & $b$ (dex) & $s$ (dex) \\
\hline
C\,{\sc i}  & $0.0116 \pm 0.0032$ & $-0.0764 \pm 0.0060$ & $0.0249 \pm 0.0056$ \\
O\,{\sc i}  & $0.0037 \pm 0.0032$ & $-0.0031 \pm 0.0068$ & $0.0320 \pm 0.0058$ \\
Na\,{\sc i} & $0.0076 \pm 0.0025$ & $-0.0493 \pm 0.0056$ & $0.0196 \pm 0.0047$ \\
Mg\,{\sc i} & $0.0078 \pm 0.0013$ & $-0.0287 \pm 0.0020$ & $0.0048 \pm 0.0038$ \\
Al\,{\sc i} & $0.0102 \pm 0.0018$ & $-0.0449 \pm 0.0024$ & $0.0075 \pm 0.0047$ \\
Si\,{\sc i} & $0.0026 \pm 0.0009$ & $-0.0105 \pm 0.0013$ & $0.0042 \pm 0.0022$ \\
S\,{\sc i}  & $0.0042 \pm 0.0024$ & $-0.0238 \pm 0.0039$ & $0.0147 \pm 0.0049$ \\
Ca\,{\sc i} & $-0.0023 \pm 0.0007$ & $0.0235 \pm 0.0013$ & $0.0035 \pm 0.0026$ \\
Sc\,{\sc i} & $0.0018 \pm 0.0012$ & $-0.0079 \pm 0.0018$ & $0.0076 \pm 0.0034$ \\
Ti\,{\sc i} & $0.0008 \pm 0.0004$ & $0.0073 \pm 0.0018$  & $0.0047 \pm 0.0021$ \\
V\,{\sc i}  & $0.0011 \pm 0.0012$ & $-0.0033 \pm 0.0021$ & $0.0071 \pm 0.0027$ \\
Cr\,{\sc i} & $-0.0018 \pm 0.0006$ & $0.0113 \pm 0.0008$ & $0.0004 \pm 0.0000$ \\
Mn\,{\sc i} & $0.0056 \pm 0.0026$ & $-0.0443 \pm 0.0061$ & $0.0229 \pm 0.0052$ \\
Co\,{\sc i} & $0.0071 \pm 0.0015$ & $-0.0440 \pm 0.0038$ & $0.0145 \pm 0.0045$ \\
Ni\,{\sc i} & $0.0043 \pm 0.0015$ & $-0.0316 \pm 0.0031$ & $0.0109 \pm 0.0027$ \\
Cu\,{\sc i} & $0.0147 \pm 0.0035$ & $-0.0854 \pm 0.0062$ & $0.0210 \pm 0.0060$ \\
Zn\,{\sc i} & $0.0076 \pm 0.0026$ & $-0.0517 \pm 0.0051$ & $0.0157 \pm 0.0047$ \\
\hline
\end{tabular}
\label{tab:myGCEabundanceagetrends}
\end{table}

\section{Bayesian analysis probability values ($p$-values)}\label{apdx:pvalues}
Tables~\ref{tab:gce_pvalues} and \ref{tab:engulfment_pvalues} present the $p$-values calculated for all 69 solar twins in our sample for the GCE and planetary engulfment models, respectively, relative to the flat and null-offset baseline models.

\begin{table}
\centering
\caption{$p$-values associated with the Bayesian evidence differences between the GCE model and the null-offset and flat baseline models.}
\label{tab:gce_pvalues}
\begin{tabular}{lcc}
\toprule
& \multicolumn{2}{c}{GCE Model} \\
\cmidrule(lr){2-3}
Star & $p_{\rm null}$ & $p_{\rm flat}$ \\
\midrule
HIP\,10175  & $<10^{-3}$ & $<10^{-3}$ \\
HIP\,101905 & $<10^{-3}$ & $<10^{-3}$ \\
HIP\,102040 & 0.001      & $<10^{-3}$ \\
HIP\,102152 & 0.009      & 0.002      \\
HIP\,10303  & 0.529      & 0.471      \\
HIP\,104045 & 0.119      & 0.496      \\
$\vdots$    & $\vdots$   & $\vdots$   \\
\bottomrule
\end{tabular}
\begin{flushleft}
\textit{Note.} The full table, including the complete solar twin sample, is available in electronic form.
\end{flushleft}
\end{table}

\begin{table*}
\centering
\caption{$p$-values associated with the Bayesian evidence differences between the engulfment models and the null-offset and flat baseline models, for the engulfment candidates identified in Section \ref{subsubsec:planet_engulfment_signatures}.}
\label{tab:engulfment_pvalues}
\resizebox{\textwidth}{!}{
\begin{tabular}{lcccccccc}
\toprule
& \multicolumn{4}{c}{GCE-corrected (B18)} & \multicolumn{4}{c}{GCE-corrected (TW)} \\
\cmidrule(lr){2-5} \cmidrule(lr){6-9}
& \multicolumn{2}{c}{Bulk Earth} & \multicolumn{2}{c}{CM Chondritic} & \multicolumn{2}{c}{Bulk Earth} & \multicolumn{2}{c}{CM Chondritic} \\
\cmidrule(lr){2-3} \cmidrule(lr){4-5} \cmidrule(lr){6-7} \cmidrule(lr){8-9}
Star & $p_{\rm null}$ & $p_{\rm flat}$ & $p_{\rm null}$ & $p_{\rm flat}$ & $p_{\rm null}$ & $p_{\rm flat}$ & $p_{\rm null}$ & $p_{\rm flat}$ \\
\midrule
HIP\,101905 & $<10^{-3}$ & $<10^{-3}$ & $<10^{-3}$ & $<10^{-3}$ & $<10^{-3}$ & $<10^{-3}$ & $<10^{-3}$ & $<10^{-3}$ \\
HIP\,77052  & $<10^{-3}$ & $<10^{-3}$ & $<10^{-3}$ & $<10^{-3}$ & $<10^{-3}$ & $<10^{-3}$ & $<10^{-3}$ & $<10^{-3}$ \\
HIP\,85042  & $<10^{-3}$ & $<10^{-3}$ & 0.001      & $<10^{-3}$ & 0.011      & 0.001      & 0.008      & 0.002      \\
HIP\,30502  & $<10^{-3}$ & $<10^{-3}$ & $<10^{-3}$ & 0.004      & 0.004      & $<10^{-3}$ & 0.024      & 0.009      \\
HIP\,40133  & 0.022      & 0.019      & 0.004      & 0.006      & 0.004      & $<10^{-3}$ & $<10^{-3}$ & $<10^{-3}$ \\
HIP\,22263  & 0.001      & $<10^{-3}$ & 0.002      & 0.001      & $<10^{-3}$ & $<10^{-3}$ & 0.004      & 0.002      \\
$\vdots$    & $\vdots$   & $\vdots$   & $\vdots$   & $\vdots$   & $\vdots$   & $\vdots$   & $\vdots$   & $\vdots$   \\
\bottomrule
\end{tabular}
}
\begin{flushleft}
\textit{Note.} The full table, including the complete solar twin sample, is available in electronic form.
\end{flushleft}
\end{table*}

\section{Engulfment analysis using GCE-corrected (TW) abundances}\label{apdx:bayesian_engulfment_myGCEcorrection}
\begin{figure}
    \centering
    \includegraphics[width=0.6\linewidth]{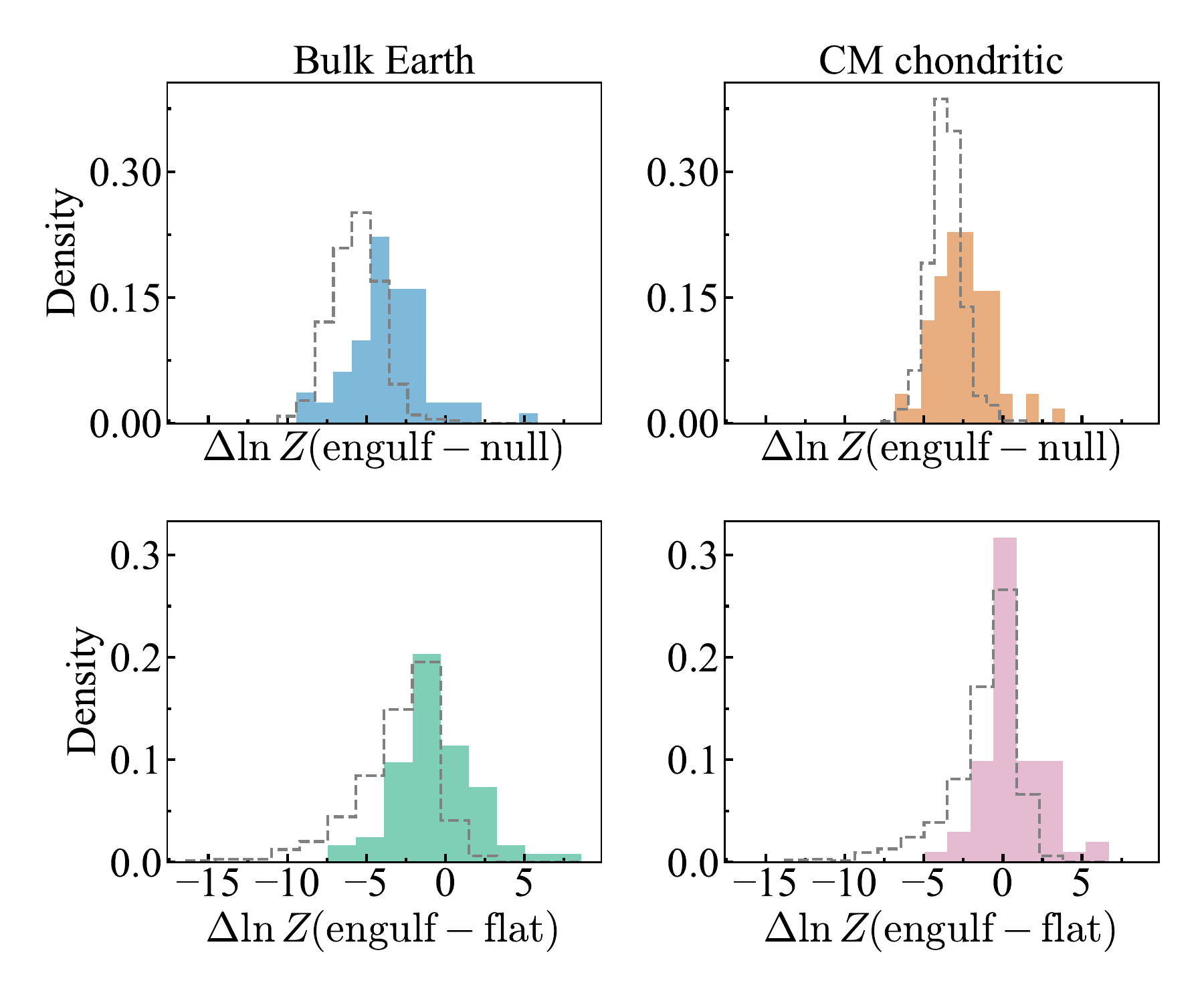}
    \caption{The same description as Figure \ref{fig:bayesian_evidence} except that the Bayesian fitting has been applied to GCE-corrected (TW) data rather than the GCE-corrected (B18) data.}
    \label{fig:engulfment_evidence_myGCEcorr}
\end{figure}

To assess the dependence of our planetary engulfment candidate identification on the GCE-correction applied, we compared the Bayesian evidence for the null-offset and flat models against that of the bulk Earth and CM chondritic engulfment models using the GCE-corrected (TW) [X/Fe] abundances (see Appendix \ref{apdx:GCEageabundancetrends}) and the corresponding mock noise datasets. The resulting $\Delta\ln Z$ distributions are shown in Figure \ref{fig:engulfment_evidence_myGCEcorr}, with final Bayesian fitting results presented in Table \ref{tab:engulfment_evidence_myGCEcorr}.

\begin{table*}
\centering
\caption{Bayesian engulfment model results using the GCE-corrected (TW) abundances, including the Bayesian evidence and best-fit parameters, for engulfment candidates identified in Section \ref{subsubsec:planet_engulfment_signatures}.}
\label{tab:engulf_results_thiswork}
\resizebox{\textwidth}{!}{
\begin{tabular}{lcccccccc}
\hline
Star &
$\Delta \ln Z_{\rm (CM-null)}$ &
$\Delta \ln Z_{\rm (CM-flat)}$ &
$M_{\rm CM}\,(M_\oplus)$ &
$\Delta \ln Z_{\rm (BE-null)}$ &
$\Delta \ln Z_{\rm (BE-flat)}$ &
$M_{\rm BE}\,(M_\oplus)$ &
$\delta_{\rm flat}$ (dex) \\
\hline
HIP\,101905 & 3.93 & 6.71 & $7.34^{+3.10}_{-2.26}$ & 5.84 & 8.62 & $5.15^{+1.41}_{-1.41}$ &
$-0.007^{+0.005}_{-0.006}$ \\
HIP\,77052 & 1.97 & 5.48 & $6.74^{+3.42}_{-2.65}$ & 1.69 & 5.20 & $3.56^{+1.34}_{-1.21}$ &
$-0.003^{+0.007}_{-0.008}$ \\
HIP\,40133 & 1.68 & 5.01 & $6.61^{+2.70}_{-2.34}$ & 0.02 & 3.35 & $2.79^{+1.01}_{-1.19}$ &
$0.001^{+0.008}_{-0.008}$ \\
HIP\,22263 & -0.02 & 3.41 & $5.12^{+3.46}_{-2.36}$ & 1.22 & 4.65 & $4.33^{+1.61}_{-1.61}$ &
$-0.007^{+0.006}_{-0.009}$ \\
HIP\,85042 & -0.24 & 3.35 & $4.94^{+3.25}_{-2.34}$ & -1.37 & 2.22 & $2.71^{+1.21}_{-1.44}$ &
$0.003^{+0.008}_{-0.007}$ \\
HIP\,30502 & -1.11 & 2.25 & $3.87^{+2.79}_{-1.86}$ & -0.43 & 2.93 & $2.80^{+1.12}_{-1.55}$ &
$-0.003^{+0.005}_{-0.006}$ \\
\vdots & \vdots & \vdots &
\vdots & \vdots & \vdots & \vdots & \vdots \\
\hline
\end{tabular}
}
\begin{flushleft}
\textit{Note.} The full table, including the complete solar twin sample, is available in electronic form.
\end{flushleft}
\label{tab:engulfment_evidence_myGCEcorr}
\end{table*}

We find that the $\Delta\ln Z$ distributions look similar to those presented in Figure \ref{fig:bayesian_evidence}, and possess similar $\Delta\ln Z$ values corresponding to the 99.6th percentile of the mock-noise distributions. Therefore we adopt the same criteria presented in Section \ref{subsubsec:planet_engulfment_signatures} to identify stars with evidence for CM chondritic or bulk Earth engulfment. Using the GCE-corrected (TW) abundances we recover two solar twins HIP\,101905 and HIP\,77052 already identified as candidates for both CM chondritic and bulk Earth engulfment based on the GCE-corrected (B18) abundances. Furthermore we identify two new solar twins exhibiting engulfment evidence, finding HIP\,40133 as a candidate for having accreted CM chondritic material, while HIP\,22263 shows evidence for ingesting bulk Earth material. Conversely, HIP\,85042 and HIP\,30502 - identified as candidates in the GCE-corrected (B18) analysis - have abundance patterns indistinguishable from the flat and null-offset models under the GCE-corrected (TW) scheme, suggesting their identification is sensitive to the choice of GCE correction.

Assuming binomial statistics and adopting the corresponding mock noise p-values (see Appendix \ref{apdx:pvalues}), the probability of recovering at least three solar twins satisfying the engulfment criteria under the baseline models is $\sim 10^{-5}$ for both the bulk Earth and CM chondritic models, supporting the interpretation that the inferred abundance patterns are physical in origin.

Finally, we acknowledge the inherent circularity in using GCE parameters derived from the full stellar sample to correct the same stellar abundances. The ideal approach is a leave-one-out scheme, in which the GCE slope, intercept, and intrinsic scatter are re-derived for each star by fitting all stars except the one being corrected, and those parameters are then used to correct that star. To test the impact of this `leave-one-out' method we implemented this approach alongside the `global-fit' method described in Appendix~\ref{apdx:GCEageabundancetrends}, and found that excluding individual stars had a negligible impact on the best-fit GCE age-abundance trend parameters. As a result, the leave-one-out approach yields nearly identical Bayesian fitting results for both the engulfment and baseline models. We therefore adopted the simpler global-fit approach for the final analysis.

\section{Bayesian fitting abundance results}\label{apdx:bayesian_engulfment_fitting}
\begin{figure}
    \centering
    \includegraphics[width=0.5\linewidth]{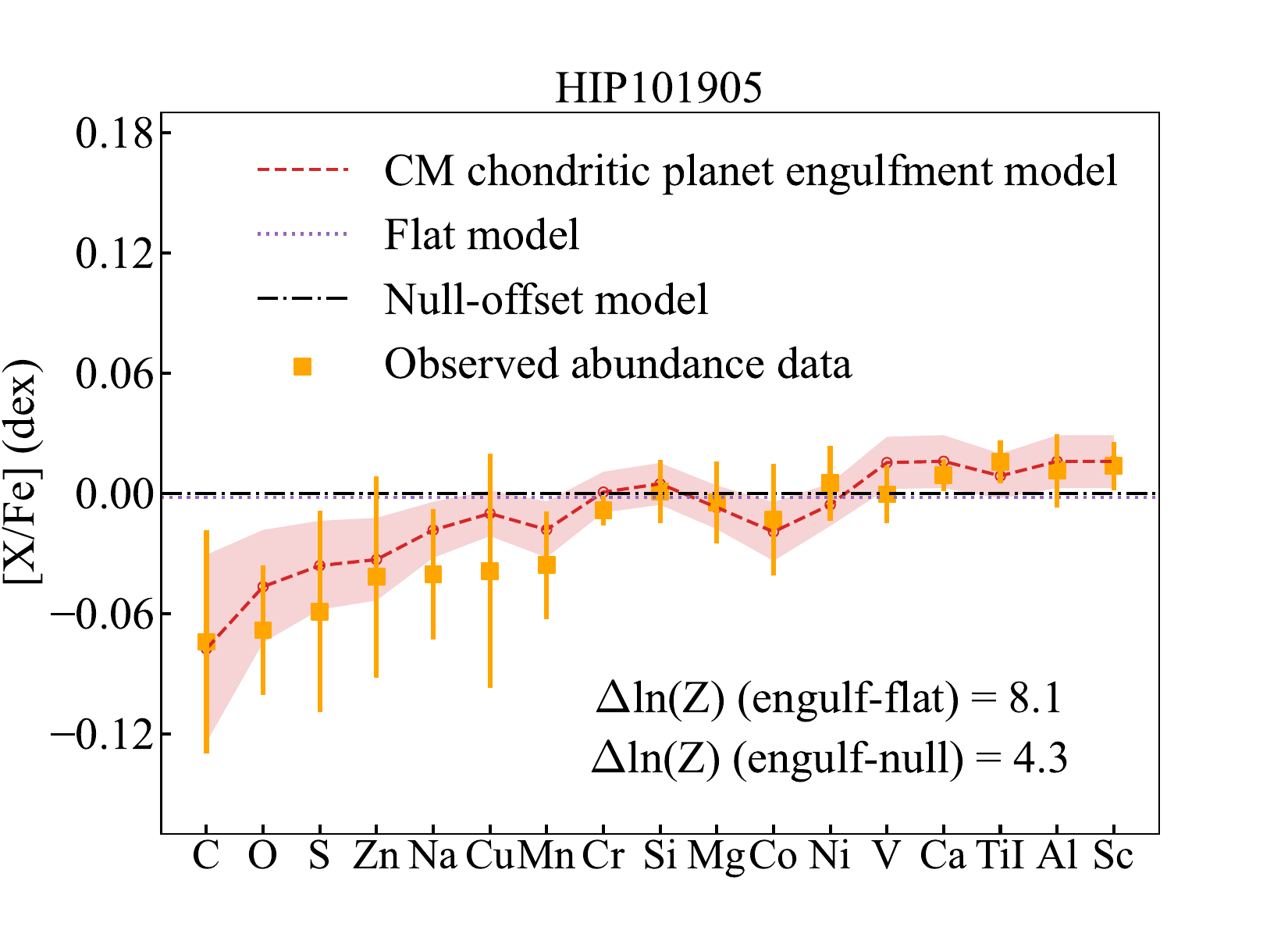}
    \caption{Bayesian abundance-pattern fits for the strongest engulfment candidate HIP\,101905, comparing the CM chondritic engulfment model to the baseline null-offset and flat models. Colours and line styles are as in Figure \ref{fig:bayesian_engulfment_abundance_patterns}.}
    \label{fig:bayesian_fitting_HIP101905_CM}
\end{figure}

\begin{figure}
    \centering
    \begin{subfigure}{0.49\textwidth}
        \centering
        \includegraphics[width=\linewidth]{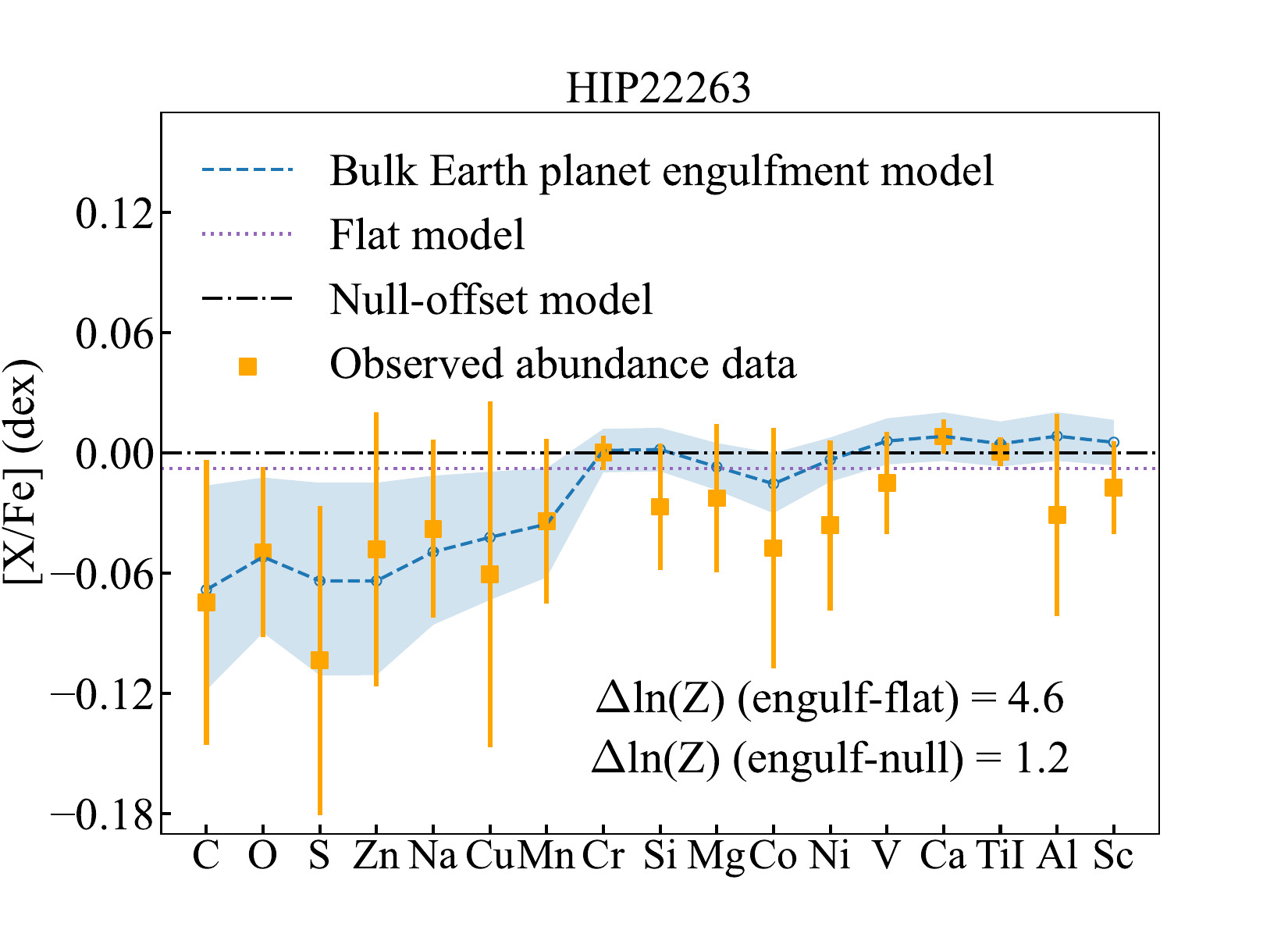}
    \end{subfigure}
    \hfill
    \begin{subfigure}{0.49\textwidth}
        \centering
        \includegraphics[width=\linewidth]{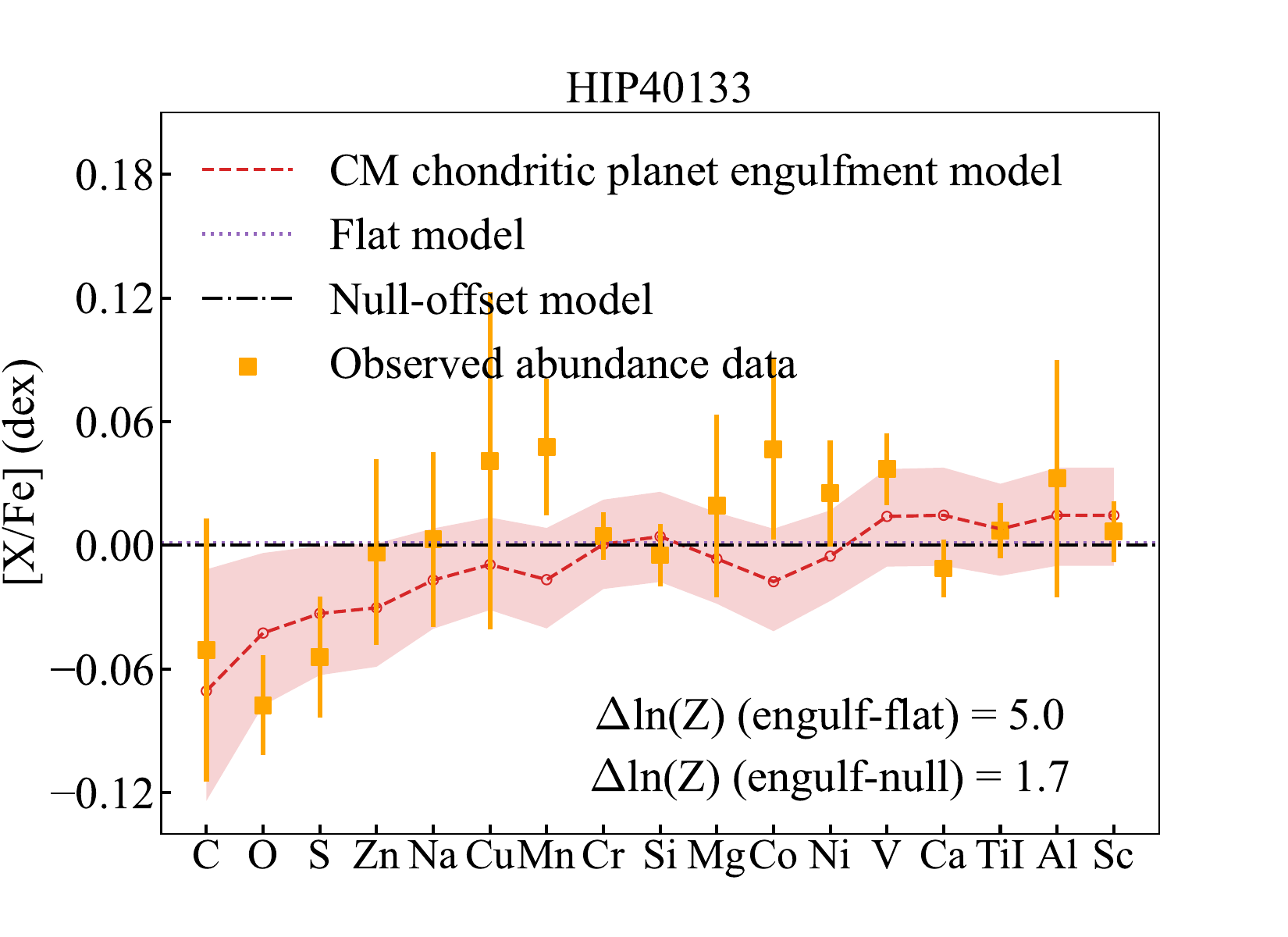}
    \end{subfigure}

    \caption{The same description as Figure \ref{fig:bayesian_engulfment_abundance_patterns}, except that the elemental abundances [X/Fe] are GCE-corrected (TW) and for tentative engulfment candidates HIP\,22263 (left; age $0.8\pm0.3$\,Gyr) and HIP\,40133 (right; age $5.4\pm0.3$\,Gyr) \citep{spina_2018}.}
    \label{fig:HIP22263_HIP40133_bayesian_abundance_patterns}
\end{figure}

\begin{figure}
    \centering

    \begin{subfigure}{0.49\textwidth}
        \includegraphics[width=\linewidth]{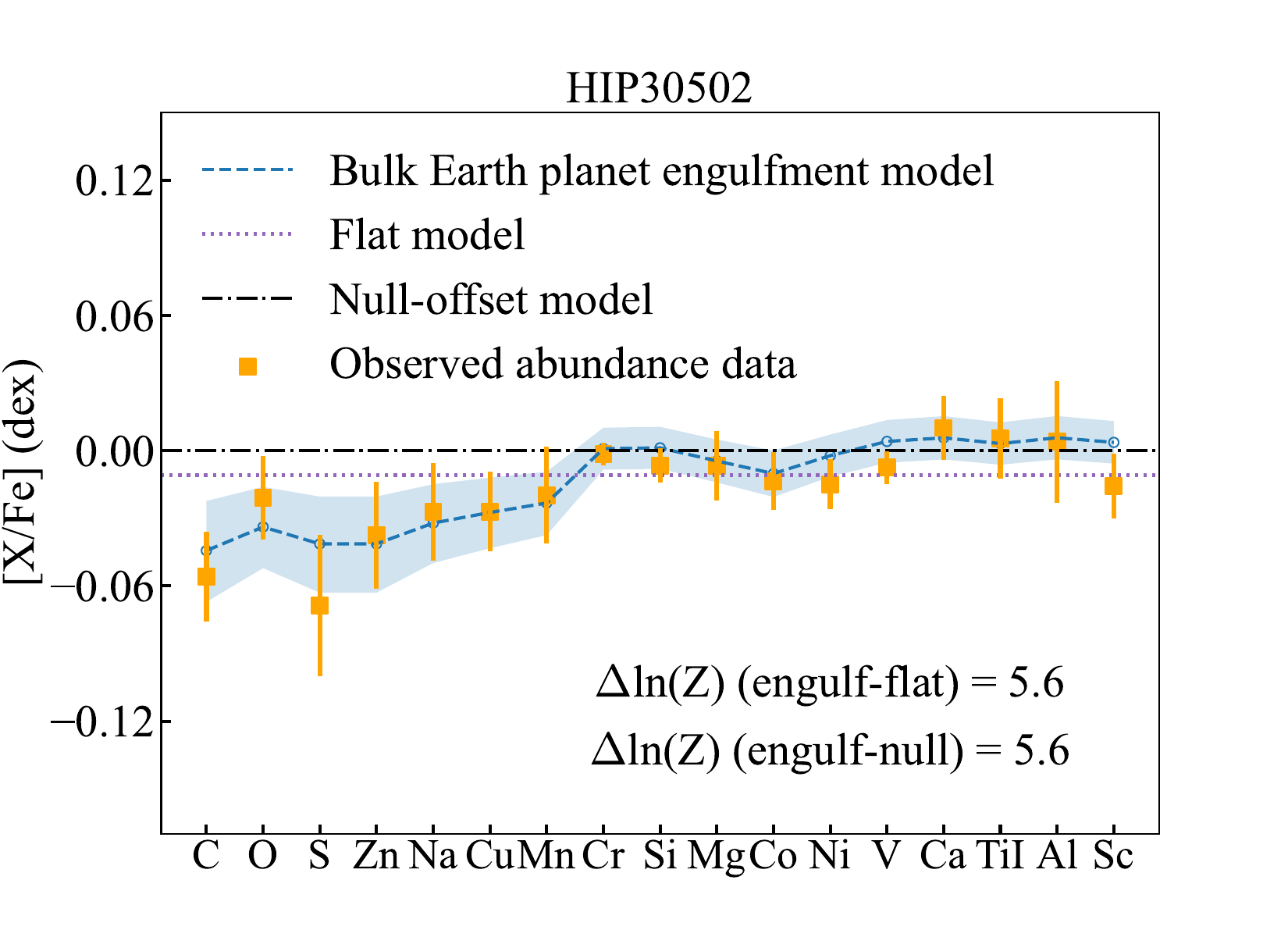}
    \end{subfigure}
    \hfill
    \begin{subfigure}{0.49\textwidth}
        \includegraphics[width=\linewidth]{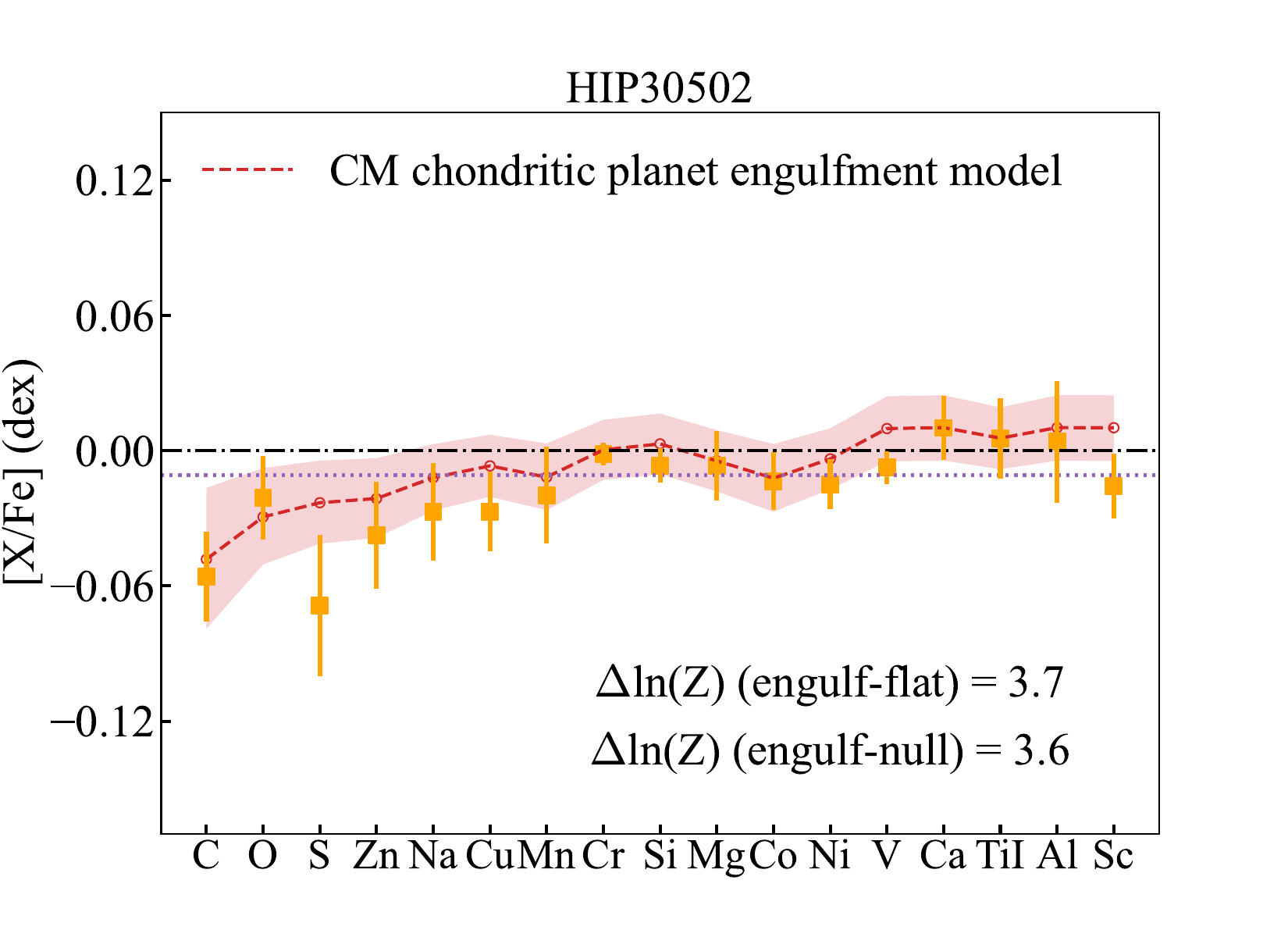}
    \end{subfigure}

    \vspace{0.5em}

    \begin{subfigure}{0.49\textwidth}
        \includegraphics[width=\linewidth]{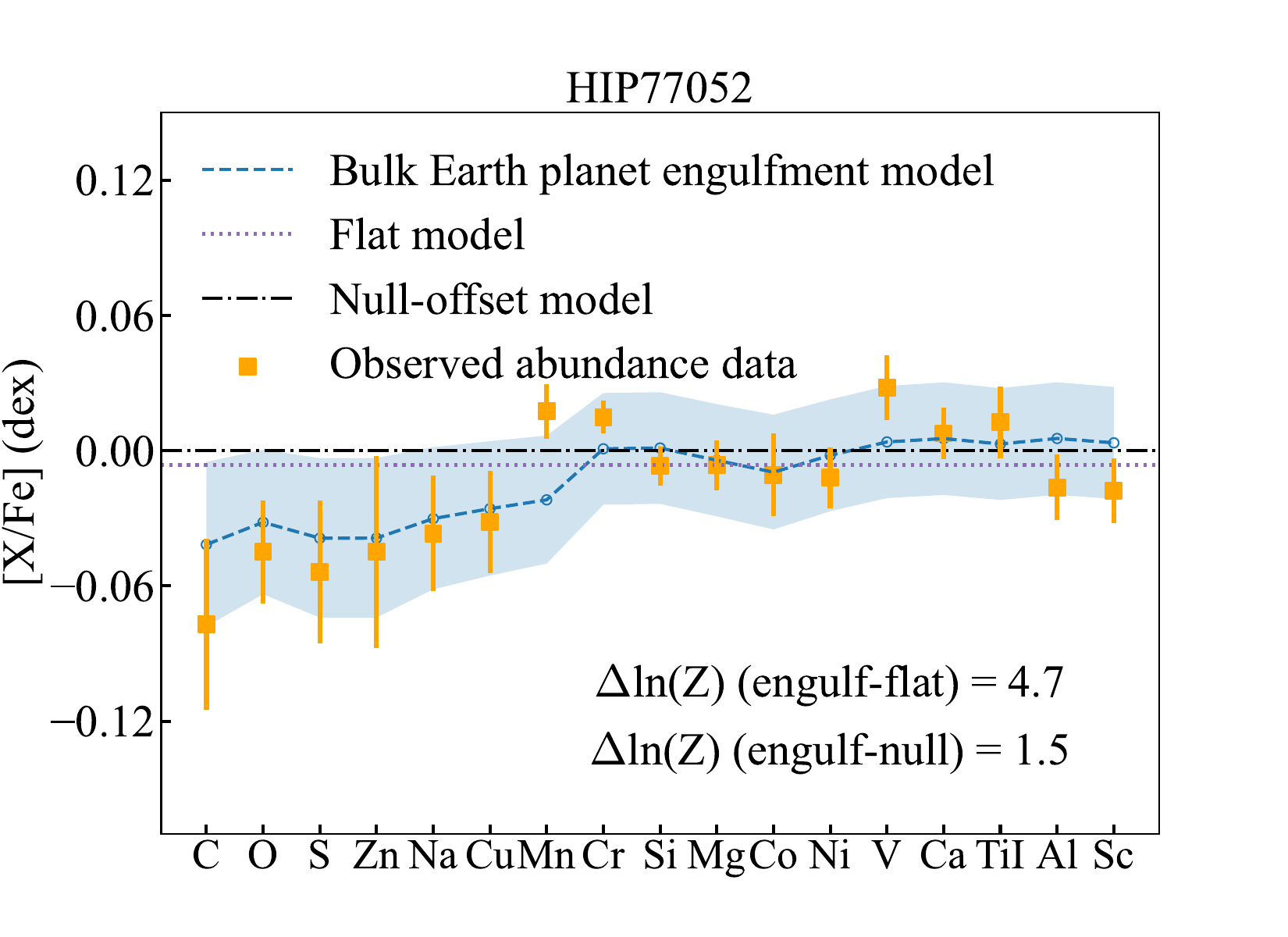}
    \end{subfigure}
    \hfill
    \begin{subfigure}{0.49\textwidth}
        \includegraphics[width=\linewidth]{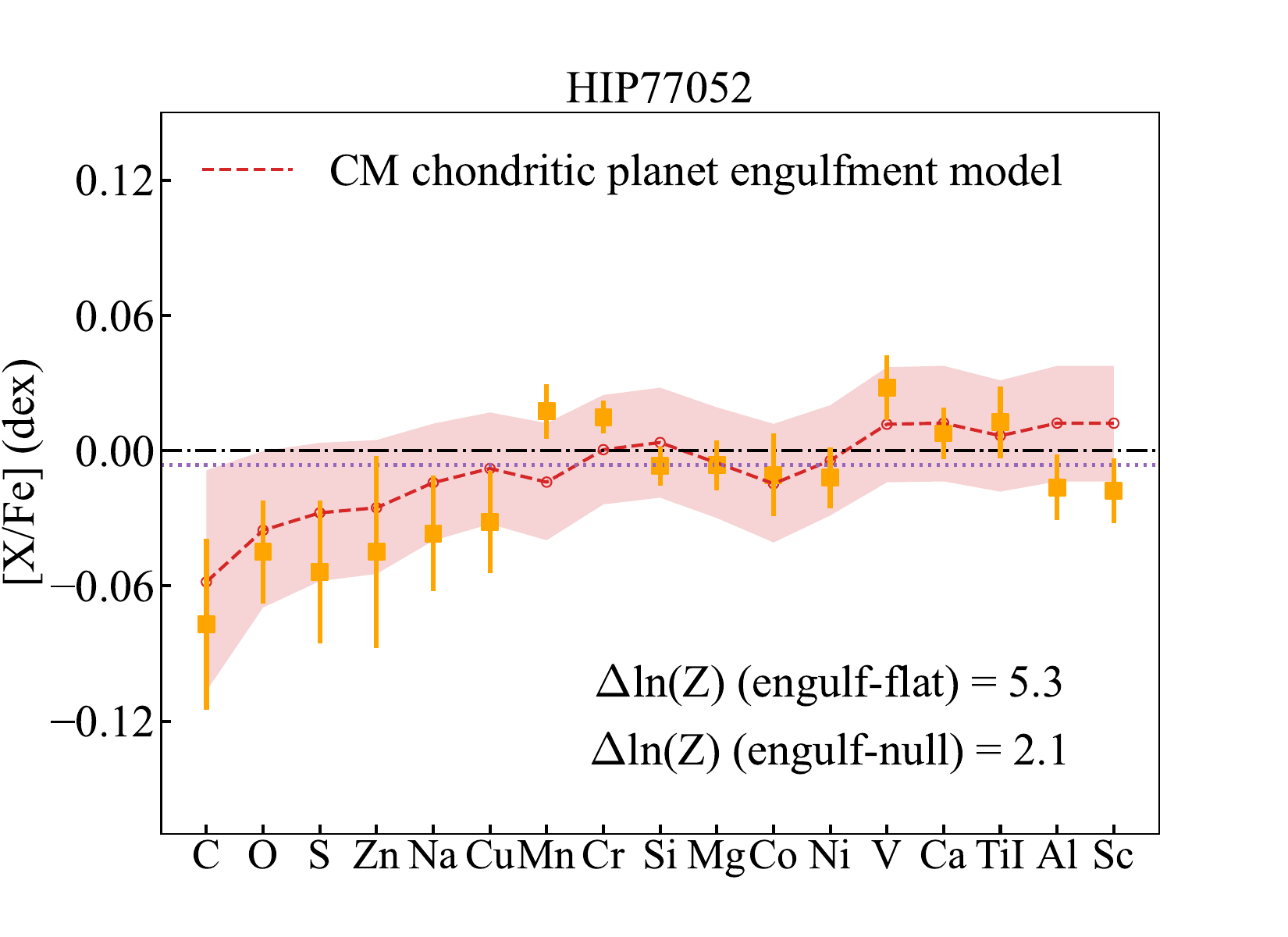}
    \end{subfigure}

    \vspace{0.5em}

    \begin{subfigure}{0.49\textwidth}
        \includegraphics[width=\linewidth]{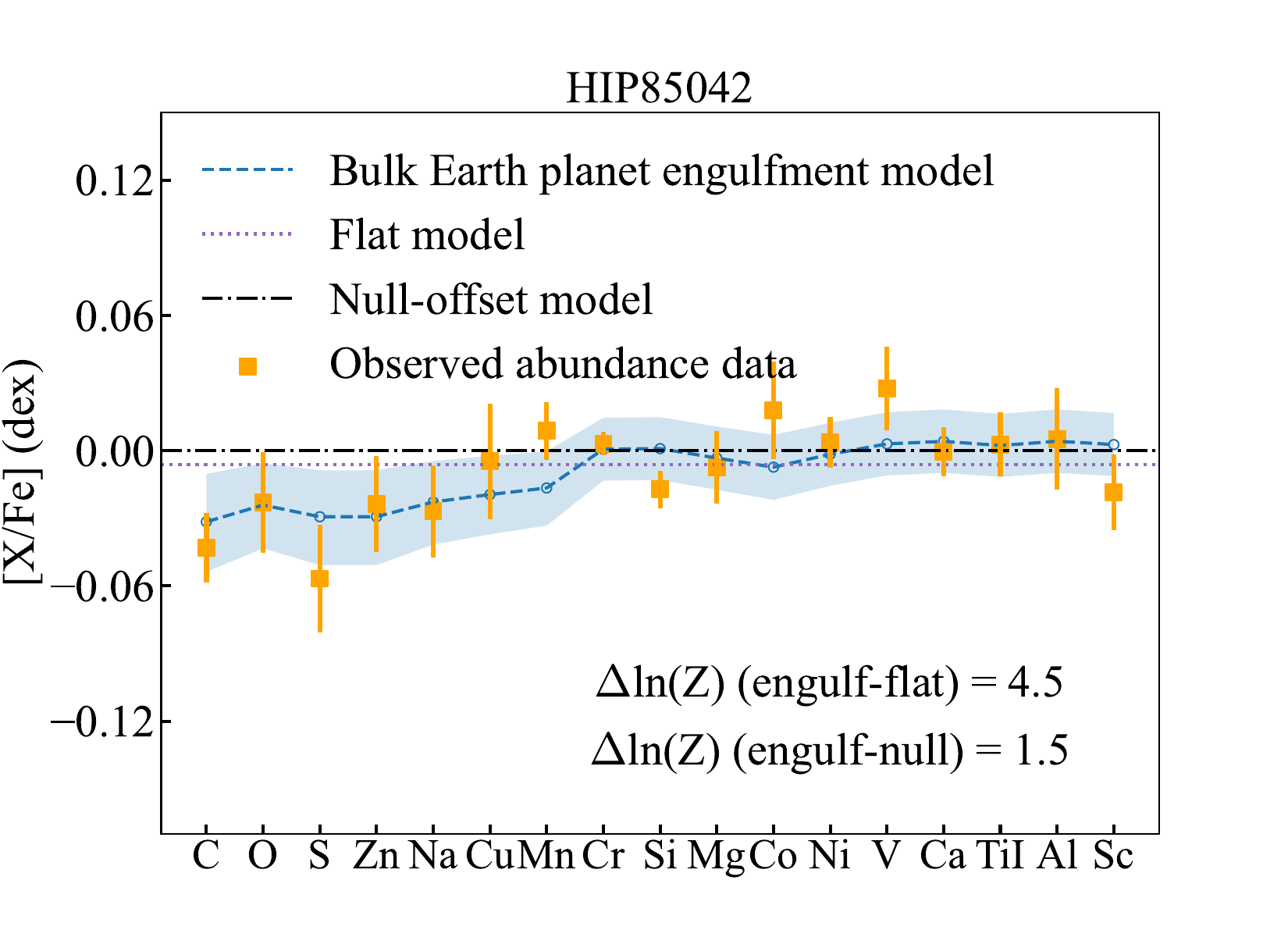}
    \end{subfigure}
    \hfill
    \begin{subfigure}{0.49\textwidth}
        \includegraphics[width=\linewidth]{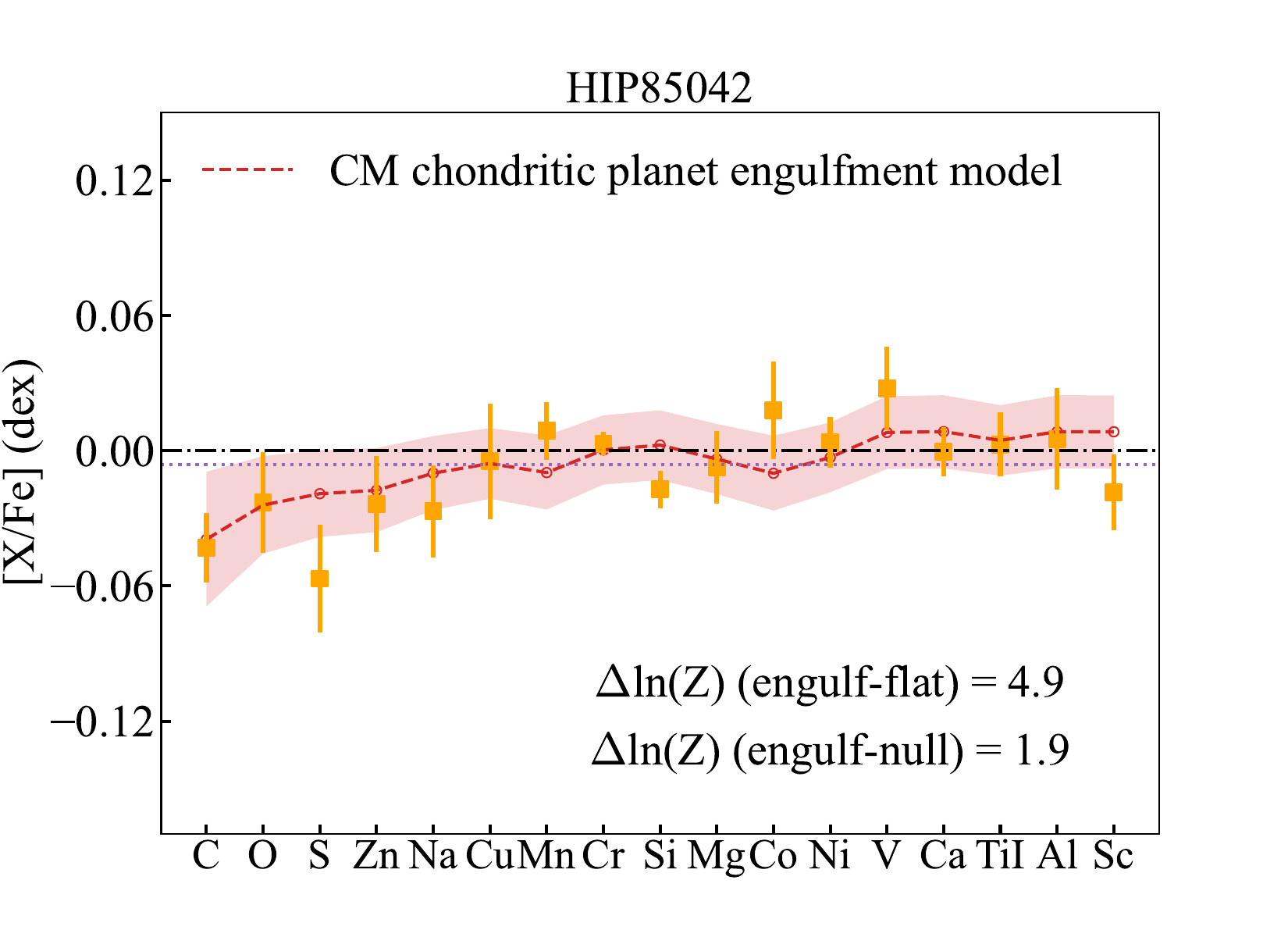}
    \end{subfigure}

    \caption{Bayesian abundance-pattern fits for the engulfment candidates HIP\,30502 (above; age 7.0$\pm$0.4\,Gyr), HIP\,77052 (middle; age 4.5$\pm$0.7\,Gyr), and HIP\,85042 (bottom; age 7.8$\pm$0.3\,Gyr) \citep{spina_2018}. For each star, the bulk Earth model is shown on the left and the CM chondritic model on the right and colours and line styles are as in Figure \ref{fig:bayesian_engulfment_abundance_patterns}.}
    \label{fig:all_candidate_bayesian_patterns}
\end{figure}

Figure \ref{fig:bayesian_fitting_HIP101905_CM} illustrates the CM chondritic engulfment model fitting results to the observed abundance data for the strongest engulfment candidate HIP\,101905. Figures \ref{fig:HIP22263_HIP40133_bayesian_abundance_patterns} and \ref{fig:all_candidate_bayesian_patterns} show the GCE-corrected abundance pattern fits for the remaining engulfment candidates identified in Section \ref{subsubsec:planet_engulfment_signatures}. Each subfigure compares the null-offset and flat baseline models against the engulfment models favoured by the Bayesian evidence; for HIP\,30502, HIP\,77052, and HIP\,85042 both bulk Earth and CM chondritic models are shown, while for HIP\,22263 and HIP\,40133 only the single favoured composition is plotted.

\end{document}